\documentstyle[12pt,epsf]{article}
\catcode`@=11
\def\chkspace{%
  \relax   
  \begingroup\ifhmode\aftergroup\dochksp@ce\fi\endgroup}
\def\dochksp@ce{%
  \unskip              
  \futurelet\chkspct@k\d@chkspc  
}
\def\d@chkspc{%
  \let\nxtsp@ce=\relax
  \ifx\chkspct@k.\else     
    \ifx\chkspct@k,\else
      \ifx\chkspct@k;\else
        \ifx\chkspct@k!\else
          \ifx\chkspct@k?\else
            \ifx\chkspct@k:\else
              \ifx\chkspct@k)\else
              \ifx\chkspct@k(\else
                \ifx\chkspct@k]\else
                  \ifx\chkspct@k-\else
                    \ifx\chkspct@k\egroup\else  
                      \let\nxtsp@ce=\put@space  
                    \fi
                  \fi
                \fi
              \fi
              \fi
            \fi
          \fi
        \fi
      \fi
    \fi
  \fi
  \nxtsp@ce
}
\def\put@space{$\;$}
\catcode`@=12

\def\ra{{$\rightarrow$}\chkspace}
\def\etal{{\it et al.}\chkspace}

\def\eg{{\it eg.}\chkspace}

\def\ep{{e$^+$e$^-$}\chkspace}
\def\epa{{e$^+$e$^-$ annihilation}\chkspace}

\def\gluino{\relax\ifmmode \tilde{g} \else $\tilde{g}$ \fi\chkspace}

\def\bbrm{\relax\ifmmode {\rm b}\bar{\rm b}
       \else ${\rm b}\bar{\rm b}$ \fi\chkspace}
\def\bb{$b\bar{b}$ \chkspace}
\def\ccrm{\relax\ifmmode {\rm c}\bar{\rm c}
       \else ${\rm c}\bar{\rm c}$ \fi\chkspace}
\def\cc{$c\bar{c}$ \chkspace}
\def\tt{\relax\ifmmode {\rm t}\bar{\rm t}
       \else ${\rm t}\bar{\rm t}$ \fi\chkspace}
\def\ss{\relax\ifmmode {\rm s}\bar{\rm s}
       \else ${\rm s}\bar{\rm s}$ \fi\chkspace}
\def\uu{\relax\ifmmode {\rm u}\bar{\rm u}
       \else ${\rm u}\bar{\rm u}$ \fi\chkspace}
\def\dd{\relax\ifmmode {\rm d}\bar{\rm d}
       \else ${\rm d}\bar{\rm d}$ \fi\chkspace}

\def\qqg{\relax\ifmmode {\rm q}\overline{\rm q}{\rm g}
\else q$\overline{\rm q}$g \fi\chkspace}

\def\bbg{$b\overline{b}g$\chkspace}

\def\afb{\relax\ifmmode A_{FB} \else
{{$A_{FB}$}}\fi\chkspace}
\def\afbb{\relax\ifmmode A_{FB}^b \else
{{$A_{FB}^b$}}\fi\chkspace}
\def\pafb{\relax\ifmmode \tilde{A}_{FB} \else
{{$\tilde{A}_{FB}$}}\fi\chkspace}
\def\pafbb{\relax\ifmmode \tilde{A}_{FB}^b \else
{{$\tilde{A}_{FB}^b$}}\fi\chkspace}

\def\pafbzo{\relax\ifmmode \tilde{A}_{FB}|_{O(0)} \else
{{$\tilde{A}_{FB}|_{O(0)}$}}\fi\chkspace}
\def\pafbfo{\relax\ifmmode \tilde{A}_{FB}|_{\oalp} \else
{{$\tilde{A}_{FB}|_{\oalp}$}}\fi\chkspace}
\def\pafbso{\relax\ifmmode \tilde{A}_{FB}|_{\oalpsq} \else
{{$\tilde{A}_{FB}|_{\oalpsq}$}}\fi\chkspace}
\def\pafbto{\relax\ifmmode \tilde{A}_{FB}|_{\oalpc} \else
{{$\tilde{A}_{FB}|_{\oalpc}$}}\fi\chkspace}

\def\pafbbzo{\relax\ifmmode \tilde{A}_{FB}^b|_{O(0)} \else
{{$\tilde{A}_{FB}^b|_{O(0)}$}}\fi\chkspace}
\def\pafbbfo{\relax\ifmmode \tilde{A}_{FB}^b|_{\oalp} \else
{{$\tilde{A}_{FB}^b|_{\oalp}$}}\fi\chkspace}
\def\pafbbso{\relax\ifmmode \tilde{A}_{FB}^b|_{\oalpsq} \else
{{$\tilde{A}_{FB}^b|_{\oalpsq}$}}\fi\chkspace}
\def\pafbbto{\relax\ifmmode \tilde{A}_{FB}^b|_{\oalpc} \else
{{$\tilde{A}_{FB}^b|_{\oalpc}$}}\fi\chkspace}

\def\afbo0{\tilde{A}_{FB}|_{O(0)}}
\def\afbo1{\tilde{A}_{FB}|_{\oalp}}
\def\afbo2{\tilde{A}_{FB}|_{\oalpsq}}
\def\afbo3{\tilde{A}_{FB}|_{\oalpc}}

\def\lam{\relax\ifmmode \Lambda_{\overline{MS}}
       \else {{$\Lambda_{\overline{MS}}$}}\fi\chkspace}
\def\lamuds{\relax\ifmmode \Lambda^{(3)}_{\overline{MS}}
       \else {{$\Lambda^{(3)}_{\overline{MS}}$}}\fi\chkspace}
\def\lamudsc{\relax\ifmmode \Lambda^{(4)}_{\overline{MS}}
       \else $\Lambda^{(4)}_{\overline{MS}}$\fi\chkspace}
\def\lamudscb{\relax\ifmmode \Lambda^{(5)}_{\overline{MS}}
       \else $\Lambda^{(5)}_{\overline{MS}}$\fi\chkspace}

\def\alp{\relax\ifmmode \alpha_s\else $\alpha_s$\fi\chkspace}
\def\alpbar{\relax\ifmmode \bar{\alpha_s}
       \else $\bar{\alpha_s}$\fi\chkspace}
\def\alpmz{\relax\ifmmode \alpha_s(M_Z)\else $\alpha_s(M_Z)$\fi\chkspace}
\def\alpmzsq{\relax\ifmmode \alpha_s(M_Z^2)
       \else $\alpha_s(M_Z^2)$\fi\chkspace}

\def\oalp{\relax\ifmmode O(\alpha_s)\else{{O($\alpha_s$)}}\fi\chkspace}
\def\oalpsq{\relax\ifmmode O(\alpha_s^2)
           \else{{O($\alpha_s^2$)}}\fi\chkspace}
\def\oalpc{\relax\ifmmode O(\alpha_s^3)
           \else{{O($\alpha_s^3$)}}\fi\chkspace}
\def\oalpf{\relax\ifmmode O(\alpha_s^4)
           \else{{O($\alpha_s^4$)}}\fi\chkspace}

\def\rb{\relax\ifmmode R_3^b/R_3^{all}
           \else{{$R_3^b/R_3^{all}$}}\fi\chkspace}
\def\rc{\relax\ifmmode R_3^c/R_3^{all}
           \else{{$R_3^c/R_3^{all}$}}\fi\chkspace}
\def\ruds{\relax\ifmmode R_3^{uds}/R_3^{all}
           \else{{$R_3^{uds}/R_3^{all}$}}\fi\chkspace}
\def\ri{\relax\ifmmode R_3^i/R_3^{all}
           \else{{$R_3^i/R_3^{all}$}}\fi\chkspace}
\def\rj{\relax\ifmmode R_3^j/R_3^{all}
           \else{{$R_3^j/R_3^{all}$}}\fi\chkspace}
\def\alpi{\relax\ifmmode \alpha^i_s/\alpha^{all}_s
           \else{{$\alpha^i_s/\alpha^{all}_s$}}\fi\chkspace}

\def\npb{Nucl. Phys.\chkspace}

\def\prl{Phys. Rev. Lett.\chkspace}
\def\prd{Phys. Rev.\chkspace}

\def\z0{{$Z^0$}\chkspace}
\def\Dst{\relax\ifmmode {\rm D}^* \else {D$^*$}\fi\chkspace}
\def\Dpl{\relax\ifmmode {\rm D}^+ \else {D$^+$}\fi\chkspace}
\def\D0{\relax\ifmmode {\rm D}^0 \else {D$^0$}\fi\chkspace}
\def\Kst{\relax\ifmmode {\rm K}^* \else {K$^*$}\fi\chkspace}
\def\K0{\relax\ifmmode {\rm K}^0_s \else {K$^0_s$}\fi\chkspace}
\def\Kpl{\relax\ifmmode {\rm K}^+ \else {K$^+$}\fi\chkspace}
\def\Kstz{\relax\ifmmode {\rm K}^{*0} \else {K$^{*0}$}\fi\chkspace}
\renewcommand{\baselinestretch}{1.5}

 1

\catcode`\@=11 
\makeatletter
\def\@seccntformat#1{\csname the#1\endcsname.\hskip 1em}

\makeatother
\begin{document}

\thispagestyle{empty}
\begin{flushright}
{\footnotesize\renewcommand{\baselinestretch}{.75}
SLAC--PUB--7660 \\
May 1998\\
}
\end{flushright}
\begin{center}
 {\Large \bf AN IMPROVED TEST OF \\
THE FLAVOR INDEPENDENCE OF STRONG  INTERACTIONS$^*$}
\vskip .3truecm
{\bf The SLD Collaboration$^{**}$}\\
Stanford Linear Accelerator Center \\
Stanford University, Stanford, CA~94309
\end{center}
 
\normalsize
 

\vskip .1truecm
 
\centerline{\bf ABSTRACT }
 
{\small
\noindent
We present an improved comparison of the strong coupling of the gluon to
light ($q_l$ = $u$+$d$+$s$), $c$, and $b$ quarks, determined from multijet 
rates in flavor-tagged samples of hadronic $Z^0$ decays recorded 
with the SLC Large Detector at the SLAC Linear Collider 
between 1993 and 1995.  Flavor separation 
among primary $q_l\overline{q_l}$, \cc, and \bb
final states was made on the basis of 
the reconstructed mass of long-lived heavy-hadron decay vertices, 
yielding tags with high purity and low bias against $\geq$ 3-jet final states. 
We find: 
$  \alpha_s^{c}/\alpha_s^{uds}=  1.036\pm0.043 \;(stat.) 
                                    ^{+0.041}_{-0.045}\;(syst.) 
                                    ^{+0.020}_{-0.018}\;(theory)$ and 
$  \alpha_s^{b}/\alpha_s^{uds} = 1.004\pm0.018\;(stat.) 
                                    ^{+0.026}_{-0.031}\;(syst.) 
                                    ^{+0.018}_{-0.029}$ $(theory)$.

}
\vskip .2truecm

\centerline{\it Submitted to Physical Review D}

\vskip .2truecm
 
\vfill

\vbox{\footnotesize\renewcommand{\baselinestretch}{1}\noindent
$^*$Work supported by Department of Energy
  contracts:
  DE-FG02-91ER40676 (BU),
  DE-FG03-91ER40618 (UCSB),
  DE-FG03-92ER40689 (UCSC),
  DE-FG03-93ER40788 (CSU),
  DE-FG02-91ER40672 (Colorado),
  DE-FG02-91ER40677 (Illinois),
  DE-AC03-76SF00098 (LBL),
  DE-FG02-92ER40715 (Massachusetts),
  DE-FC02-94ER40818 (MIT),
  DE-FG03-96ER40969 (Oregon),
  DE-AC03-76SF00515 (SLAC),
  DE-FG05-91ER40627 (Tennessee),
  DE-FG02-95ER40896 (Wisconsin),
  DE-FG02-92ER40704 (Yale);
  National Science Foundation grants:
  PHY-91-13428 (UCSC),
  PHY-89-21320 (Columbia),
  PHY-92-04239 (Cincinnati),
  PHY-95-10439 (Rutgers),
  PHY-88-19316 (Vanderbilt),
  PHY-92-03212 (Washington);
  The UK Particle Physics and Astronomy Research Council
  (Brunel, Oxford and RAL);
  The Istituto Nazionale di Fisica Nucleare of Italy
  (Bologna, Ferrara, Frascati, Pisa, Padova, Perugia);
  The Japan-US Cooperative Research Project on High Energy Physics
  (Nagoya, Tohoku);
  The Korea Research Foundation (Soongsil, 1997).}

\vfill
\eject

\noindent {\Large \bf 1. Introduction} 
  
\vskip .5truecm

\noindent
In order for Quantum Chromodynamics (QCD)~\cite{qcd} to be a gauge-invariant
renormalisable field theory it is required that 
the strong coupling between quarks ($q$) and gluons ($g$), $\alpha_s$, 
be independent of quark flavor.
This basic {\it ansatz} can be tested directly in \epa by 
measuring the strong coupling in
events of the type \ep \ra $q{\bar q}g$ for specific quark flavors.
Whereas an absolute determination of $\alpha_s$ using such a technique is
limited, primarily by large theoretical uncertainties, to the 5\%-level of
precision~\cite{phil}, a  much more precise test of the flavor-independence
can be made from the ratio of the couplings for different quark flavors,
in which most experimental errors and theoretical uncertainties cancel. 
Furthermore,  the emission of
gluon radiation in \bb events is expected~\cite{bmass} 
to be modified relative to that
in $q_l\overline{q_l}$ ($q_l$ =$u$+$d$+$s$) 
events due to the large $b$-quark mass, and  
comparison of the rates for \z0 \ra \bbg and \z0 \ra $q_l\overline{q_l}g$ may 
allow measurement of the running mass\footnote{Use of the
modified minimal subtraction renormalisation scheme~\cite{msbar}
is implied throughout this paper.} 
of the $b$-quark, $m_b(M_{Z^0})$\footnote{The DELPHI
Collaboration has recently measured the three-jet rate ratio $R_3^b/R_3^{uds}$
to a precision of $\pm0.009$, and, under the assumption of a
flavor-independent strong coupling, derived a value of the running
$b$-mass~\cite{DELPHI97}; this issue will be discussed in Section 6.}.
Finally, in addition to providing a powerful test of QCD, such measurements
allow constraints to be placed on physics beyond the Standard Model.
For example, a flavor-dependent anomalous quark chromomagnetic 
moment would modify~\cite{rizzo} the emission rate of gluons for the different
quark flavors, and would manifest itself in the form of an apparently 
flavor-dependent strong coupling. 

The first such comparisons, 
of $\alpha_s$ for $c$ or $b$ quarks with $\alpha_s$ for all flavors,
were made at the PETRA \ep collider at c.m. energies in the range
$35\leq \sqrt{s}\leq 47$ GeV and were limited in precision to 
$\delta\alpha_s^c/\alpha_s^{all}$ = 0.41 and 
$\delta\alpha_s^b/\alpha_s^{all}$ = 0.57~\cite{TASSO}                
due to the small data sample and limited heavy-quark tagging capability.
These studies made the 
simplifying assumptions that $\alpha_s^b$ = $\alpha_s^{uds}$ and
$\alpha_s^c$ = $\alpha_s^{uds}$, respectively. More recently,
measurements made at the \z0 resonance have benefitted from the use of
micro-vertex detectors for improved heavy-quark tagging. 
Samples of tagged \bb events recorded at LEP have been used to test
flavor-independence to a precision of $\delta\alpha^b_s/\alpha_s^{all}$
= 0.012~\cite{LEPalphas,aleph}, but these measurements
were insensitive to any differences among $\alpha_s$ values for the 
non-$b$-quarks.
The ALEPH Collaboration also measured $\alpha_s^{bc}/\alpha_s^{uds}$
to a precision of $\pm0.023$~\cite{aleph}, but in this case there is no
sensitivity to a different \alp for $c$ and $b$ quarks.

The OPAL Collaboration has measured
$\alpha^f_s/\alpha_s^{all}$  for all five flavors $f$ with no assumption 
on the relative value of $\alpha_s$ for the different flavors~\cite{OPAL},
and has verified flavor-independence to a
precision of 
$\delta \alpha^b_s/\alpha_s^{all}$ = 0.026, 
$\delta \alpha^c_s/\alpha_s^{all}$ = 0.09, 
$\delta \alpha^s_s/\alpha_s^{all}$ = 0.15, 
$\delta \alpha^d_s/\alpha_s^{all}$ = 0.20, and
$\delta \alpha^u_s/\alpha_s^{all}$ = 0.21. 
In that analysis the precision of the test was limited by 
the kinematic signatures used to tag $c$ and light-quark events, which suffer
from low efficiency and strong biases against events containing hard gluon 
radiation. In our previous study~\cite{MHalphas} we used hadron lifetime
information as a basis for separation of \bb, \cc and
light-quark events with relatively small bias against 3-jet final states.
We verified flavor-independence to a precision of 
$\delta \alpha^b_s/\alpha_s^{all}$ = 0.06, 
$\delta \alpha^c_s/\alpha_s^{all}$ = 0.17, and 
$\delta \alpha^s_{uds}/\alpha_s^{all}$ = 0.04.
 
Here we present an improved test of the flavor-independence of strong
interactions using a sample of hadronic \z0 decay
events produced by the SLAC Linear Collider (SLC) and recorded in the
SLC Large Detector (SLD) in data-taking runs between 1993 and 1995.
The precise tracking capability of the Central Drift Chamber
and the 120-million-pixel CCD-based Vertex Detector (VXD2), 
combined with the stable, micron-sized beam interaction point (IP), allowed
us to reconstruct topologically secondary vertices from heavy-hadron decays
with high efficiency. High-purity samples of \z0 \ra \bb(g) and \z0 \ra \cc(g)
events were then tagged on the basis of the reconstructed mass and momentum of 
the secondary vertex. Events containing no secondary vertex and no tracks 
significantly displaced from the IP were tagged as a high-purity 
$Z^0 \rightarrow q_l \bar{q_l}(g)$ event sample.
The method makes no assumptions about the relative values         
of $\alpha^{b}_s$, $\alpha^{c}_s$ and $\alpha^{uds}_s$.
Furthermore, an important advantage of the method is that it has  
low bias against $\geq3$-jet events. In addition to using an
improved flavor-tagging technique, 
this analysis utilises a data sample three times larger than that used for
our previous measurement, and allows us to test 
the flavor independence of 
strong interactions to a precision higher by roughly a factor of three.
Finally, quark mass effects in \z0 \ra $q\bar{q}g$ events have recently been
calculated~\cite{Aachen,Rodrigo} at next-to-leading order in perturbative QCD,
and are non-negligible on the scale of our experimental errors; 
we have utilised these calculations in this analysis.

\vskip 1truecm

\noindent{\Large \bf 2. Apparatus and Hadronic Event Selection} 
 
\vskip .5truecm

\noindent
This analysis is based on roughly 150,000 hadronic events produced in 
\ep annihilations 
at a mean center-of-mass energy of $\sqrt{s}=91.28$ GeV.
A general description of the SLD can be found elsewhere~\cite{sld}.
The trigger and initial selection criteria for hadronic $Z^0$ decays are 
described in Ref.~\cite{sldalphas}.
This analysis used charged tracks measured in the Central Drift
Chamber (CDC)~\cite{cdc} and in the Vertex Detector (VXD2)~\cite{vxd}.
Momentum measurement is provided by a uniform axial magnetic field of 0.6T.
The CDC and VXD2  give a momentum resolution of
$\sigma_{p_{\perp}}/p_{\perp}$ = $0.01 \oplus 0.0026p_{\perp}$,
where $p_{\perp}$ is the track momentum transverse to the beam axis in
GeV/$c$. In the plane normal to the beamline 
the centroid of the micron-sized SLC IP was reconstructed from tracks
in sets of approximately thirty sequential hadronic \z0 decays to a precision 
of $\sigma_{IP}\simeq7$ $\mu$m. 
Including the uncertainty on the IP position, the resolution on the 
charged-track impact parameter ($d$) projected in the plane perpendicular
to the beamline is $\sigma_d$ =
11$\oplus$70/$(p_{\perp} \sin^{3/2}\theta)$ $\mu$m, where
$\theta$ is the track polar angle with respect to the beamline. 
The event thrust axis~\cite{thrust} was calculated using energy clusters
measured in the Liquid Argon Calorimeter~\cite{lac}. 

A set of cuts was applied to the data to select well-measured tracks
and events well contained within the detector acceptance.
Charged tracks were required to have a distance of
closest approach transverse to the beam axis within 5 cm,
and within 10 cm along the axis from the measured IP,
as well as $|\cos \theta |< 0.80$, and $p_\perp > 0.15$ GeV/c.
Events were required to have a minimum of seven such tracks,
a thrust axis  polar angle w.r.t. the beamline, $\theta_T$,
within $|\cos\theta_T|<0.71$, and
a charged visible energy $E_{vis}$ of at least 20~GeV,
which was calculated from the selected tracks assigned the charged pion mass. 
The efficiency for selecting a well-contained $Z^0 \rightarrow q{\bar q}(g)$
event was estimated to be above 96\% independent of quark flavor. The
selected sample comprised 77,896 events, with an estimated
$0.10 \pm 0.05\%$ background contribution dominated
by $Z^0 \rightarrow \tau^+\tau^-$ events.
 
For the purpose of estimating the efficiency and purity of the event
flavor-tagging procedure we made use of a detailed Monte Carlo (MC) simulation 
of the detector. The
JETSET 7.4~\cite{jetset} event generator was used, with parameter
values tuned to hadronic \ep annihilation data~\cite{tune},
combined with a simulation of $B$-hadron decays
tuned~\cite{sldsim} to $\Upsilon(4S)$ data and a simulation of the SLD
based on GEANT 3.21~\cite{geant}.
Inclusive distributions of single-particle and event-topology observables
in hadronic events were found to be well described by the
simulation~\cite{sldalphas}. Uncertainties in the simulation 
were taken into account in the systematic errors (Section 5). 

\vskip 1truecm

\noindent{\Large \bf 3. Flavor Tagging} 
 
\vskip .5truecm

\noindent
Separation of the accepted event sample into tagged flavor subsamples was
based on the invariant mass of topologically-reconstructed  
long-lived heavy-hadron decay
vertices, as well as on charged-track impact parameters
in the plane normal to the beamline. 
In each event a jet structure was defined as a basis for 
flavor-tagging by applying the `JADE' jet-finding algorithm~\cite{jade}
to the selected tracks; a value of the normalised jet-jet invariant-mass
parameter $y_c$ = 0.02 was used. 
The impact parameter of each track, $d$, was given a positive (negative) sign 
according to
whether the point-of-closest approach to its jet axis was on the same side
(opposite side) of the IP as the jet.
Charged tracks used for the subsequent event flavor-tagging
were further required to have at least 40 hits in the CDC, with the first
hit at a radial distance of less than 39 cm from the beamline, at least 
one VXD2 hit, a combined CDC + VXD2 track fit quality of $\chi^2_{dof}$ 
$<$ 5, momentum $p$ $>$ 0.5 GeV/$c$, a distance of
closest approach transverse to the beam axis within 0.3 cm,
and within 1.5 cm along the axis from the measured IP, and
an error on the impact parameter, $\sigma_d$, less than 250$\mu$m. 
Tracks from identified K$_s^0$ and $\Lambda$ decays and $\gamma$ conversions
were removed.

In each jet we then searched for a secondary 
vertex (SV), namely a vertex spatially separated from the measured IP.
In the search those tracks were considered that were  assigned to the
jet by the jet-finder.
Individual track probability-density functions in 3-dimensional co-ordinate 
space were examined and 
a candidate SV was defined by a region of high track overlap density; the
method is described in detail in~\cite{jackson}. 
A SV was required to contain two or more tracks, 
and to be separated from the IP by at least 1~mm.
We found 14,096 events containing a SV in only one jet, 5817 events containing
a SV in two jets, and 54 events containing a SV in more than two jets.
The selected SVs comprise, on average, 3.0 tracks.
These requirements preferentially select SVs that originate from the decay 
of particles with relatively long lifetime.
In our simulated event sample a SV was found in 50\% of all true $b$-quark
hemispheres, in 15\% of true $c$-quark, and in $<1\%$ of
true light-quark hemispheres~\cite{jackson}, where hemispheres were defined by
the plane normal to the thrust axis that contains the IP.

Due to the cascade structure of $B$-hadron decays, not all the tracks
in the decay chain will necessarily originate from a common decay point, 
and in such cases the SV may not be fully reconstructed in \bb events.
Therefore, we improved our estimate of the SV by allowing the possibility of
attaching additional tracks. 
First, we defined the vertex axis to be the straight line joining
the IP and the SV centroids, and $D$ to be the distance along this axis
between the IP and the SV. For each track in the jet not included in the SV the 
point of closest approach (POCA), and corresponding distance of closest 
approach, $T$, to the vertex axis were determined. 
The length, $L$, of the projection of the vector joining the IP and the POCA, 
along the vertex axis was then calculated. 
Tracks with $T<1.0$~mm, $L>0.8$~mm and $L/D>0.22$ were then attached 
to the SV.
On average 0.5 tracks per SV were attached in this fashion.

The invariant mass, $M_{ch}$, of each SV was then calculated by assigning each 
track the charged pion mass. 
In order to account partially for the effect of neutral particles missing
from the SV we applied a kinematic correction to the calculated $M_{ch}$.
We added the momentum vectors of all tracks forming the SV to obtain the
vertex momentum, $\vec{P_{vtx}}$, and evaluated the
magnitude of the component of the vertex momentum tranverse to the vertex 
axis, $P_t$. In order to reduce the effect of the IP and SV measurement errors,
the vertex axis was varied 
within an envelope defined by all possible cotangents to the 
error ellipsoids of both the IP and the SV, and the minimum $P_t$ was
chosen. We then defined the $P_t$-corrected vertex mass, 
$M_{vtx} = \sqrt{M_{ch}^{2}+P_{t}^{2}} + |P_{t}|$. 

The distributions of $M_{vtx}$ and $P_{vtx}$ are shown in Fig.~1; the data are
reproduced by the simulation, in which the primary event-flavor
breakdown is indicated. The region $M_{vtx}$ $>$ 2
GeV/$c^2$ is populated predominantly by \z0 \ra \bb events, whereas the region
$M_{vtx}$ $<$ 2 GeV/$c^2$ is populated roughly equally by \bb and non-\bb 
events. 

In order to optimise the separation among flavors we examined the 
two-dimensional distribution of $P_{vtx}$ vs. $M_{vtx}$.
The distribution for events containing a SV is
shown in Fig.~2 for the data and simulated samples; 
the data (Fig.~2a) are reproduced by the simulation (Fig.~2b).
The distributions for the simulated subsamples corresponding to true primary 
\bb, \cc, and $q_l\overline{q_l}$ events are shown in Figs.~2c, 2d and 2e
respectively. 

In order to separate \bb and \cc events from each other, and from the 
$q_l\overline{q_l}$ events, we defined the regions:
(A) $M_{vtx}>1.8$ $\oplus$ $P_{vtx}+10<15M_{vtx}$; 
(B) $M_{vtx}<1.8$ $\oplus$ $P_{vtx}>5$ $\oplus$ $P_{vtx}+10\geq15M_{vtx}$; 
where $M_{vtx}$ ($P_{vtx}$) is in units of GeV/$c^2$ (GeV/$c$); 
(C) all remaining events containing a SV. The boundaries of
regions (A) and (B) are indicated in
Figs. 2c and 2d, respectively, and all three regions are labelled in Fig.~2f. 
The $b$-tagged sample (subsample~1) was defined to comprise those
events containing any vertex in region (A).  
For the remaining events containing any vertex in region (B) we 
examined the distribution of the
impact parameter of the vector $\vec{P_{vtx}}$ w.r.t. the IP, $\delta_{vtx}$ 
(Fig.~3); according to the simulation true primary 
\cc events dominate the population in the region $\delta_{vtx}<0.02$ cm.
Therefore, we defined the $c$-tagged sample (subsample 2) to comprise 
those events in region (B) with $\delta_{vtx}<0.02$ cm. 

Events containing no selected SV were then examined. 
For such events the distribution of $N_{sig}$,
the number of tracks per event that miss the IP by $d>2\sigma_d$,
is shown in Fig.~4. The $uds$-tagged sample (subsample 3) was defined to 
comprise those events with $N_{sig}$ = 0.  
All events not assigned to subsamples 1,2 or 3 were defined to comprise the 
untagged sample (subsample 4).
Using the simulation we estimated that the efficiencies
$\varepsilon^{ji}$ for selecting events (after acceptance cuts) of type 
$i$ ($i= b,\,c,\,uds,\,$) into subsample $j$ ($1\leq j\leq 4$), and 
the fractions $\Pi^{ji}$ of events of type $i$ in subsample $j$, are
$(\varepsilon,\Pi)^{1\;b} = (61.5\pm 0.1\%, 95.5 \pm 0.1\%)$,  
$(\varepsilon,\,\Pi)^{2\;c} = (19.1\pm0.1\%,\, 64.4\pm 0.3\%)$ and 
$(\varepsilon,\,\Pi)^{3\;uds} = (56.4\pm 0.1\%,\, 90.6\pm 0.1\%)$.
The composition of the untagged sample (subsample 4) was estimated to be
$\Pi^{4\;uds} = 59.3 \pm 0.1$\%, 
$\Pi^{4\;c} = 24.1 \pm 0.1$\% and
$\Pi^{4\;b} = 16.6 \pm 0.1$\%. 
The errors on these values are discussed in Section~5.          

\vskip 1truecm

\noindent{\Large \bf 4. Jet Finding}

\vskip .5truecm

\noindent
For the study of flavor-independence the jet structure of events 
was reconstructed in turn using six iterative clustering algorithms.
We used the `E', `E0', `P', and `P0' variations of the JADE algorithm, 
as well as the `Durham' (`D') and `Geneva' (`G') algorithms~\cite{Siggi}.
In each case events were divided into two categories: those containing (i) two
jets, and (ii) three or more jets. The fraction of the event sample in
category (ii) was defined as the 3-jet rate $R_3$.  
This quantity is infrared- and collinear-safe
and has been calculated to ${\cal O}(\alpha_s^2)$ in perturbative 
QCD~\cite{Siggi,KN}. For each algorithm we repeated the subsequent 
analysis successively across a
range of values of the normalised jet-jet invariant-mass parameter $y_c$,
$0.005\leq y_c\leq 0.12$.
The ensemble of results from the different $y_c$
values was used to cross-check the consistency of the method.
In the final stage an `optimal' $y_c$ value was chosen for each algorithm 
so as to minimise the overall error on the analysis, and the spread in
results over the algorithms was used to assign an additional uncertainty
(Section 7).

Each of the six jet-finding algorithms was applied to
each tagged-event subsample $j$, $1\leq j\leq 3$ (Section 3), as well as to the
global sample of all accepted events (`$all$'). 
For each algorithm the 3-jet rate in each subsample 
was calculated,
and the ratios $R_3^j/R_3^{all}$, in which many systematic errors
should cancel, were then derived. As an
example the $R_3^j/R_3^{all}$ are shown as a function of $y_c$ for the JADE 
E0 algorithm in Fig.~5a. The results of the corresponding
analysis applied to the simulated event sample are also shown; the simulation 
reproduces the data. Similar results were obtained for the other jet
algorithms (not shown).
 
For each algorithm and $y_c$ value
the $R_3^i$ for each of the $i$ quark types ($i= b,\,c,\,uds$)
was extracted from a simultaneous maximum likelihood fit to 
 $n_2^j$ and $n_3^j$,
 the number of 2-jet and 3-jet events, respectively,
in the flavor-tagged subsample ($1\leq j \leq 3$), using the relations:
\begin{eqnarray}  
&  n_{2}^j\ =&\ \sum_{i=uds,c,b} \left(\varepsilon_{(2\rightarrow 2)}^{ji}
 (1-R_3^i) + \varepsilon_{(3\rightarrow 2)}^{ji} R_3^i\right) f^i N  
\nonumber  \\
& n_{3}^j\ =&\ \sum_{i=uds,c,b} \
\left( \varepsilon_{(3\rightarrow 3)}^{ji}
 R_3^i +  \varepsilon_{(2\rightarrow 3)}^{ji} (1-R_3^i)\right) f^i N \ . 
 \end{eqnarray} 
Here $N$ is the total number of events after correction 
for the event selection efficiency, and 
$f^i$ is the Standard Model fractional hadronic width for $Z^0$ 
decays to quark type~$i$. The $y_c$-dependent 
$3\times3$ matrices $\varepsilon_{(2\rightarrow 2)}^{ji}$ and 
$\varepsilon_{(3\rightarrow 3)}^{ji}$ are the efficiencies
for an event of type $i$, with  2- or 3-jets at the parton level, to pass all 
cuts and enter subsample {\it j} as a 2- or 3-jet event, respectively.
Similarly, the $3\times3$ matrices $\varepsilon_{(2\rightarrow 3)}^{ji}$ and 
$\varepsilon_{(3\rightarrow 2)}^{ji}$ are the efficiencies
for an event of type $i$, with  2- or 3-jets at the parton level, to pass all 
cuts and enter subsample {\it j} as a 3- or 2-jet event, respectively.
These matrices were calculated from the Monte Carlo simulation, and the
systematic errors on the values of the matrix elements are discussed in 
Sections 5 and 6. 

This formalism explicitly accounts for modifications of the
parton-level 3-jet rate due to hadronisation, detector effects, 
and flavor-tagging bias. The latter effect is evident, for the E0 algorithm, 
in Fig.~5a,
where it can be seen that the measured values of $R_3^j/R_3^{all}$ are below
unity for subsamples $j$ = 1,2 and 3, implying that the flavor tags
preferentially select 2-jet rather than 3-jet events. 
 For example, at $y_c$ = 0.02 the normalised difference in
efficiencies for correctly tagging a 2-jet event and a 3-jet event of type $i$ 
in subsample $j$ are ${\cal B}^{1,b}$=5.7\%,
${\cal B}^{2,c}$=14.5\%, and ${\cal B}^{3,uds}$=4.1\%, where 
${\cal B}^{ji}\equiv(\varepsilon^{ji}_{2 \rightarrow 2} - 
\varepsilon^{ji}_{3\rightarrow 3})/\varepsilon^{ji}_{2\rightarrow 2}$; 
these biases are considerably smaller than those found in~\cite{OPAL}, which 
resulted from the kinematic signatures employed for flavor-tagging.
It should be noted that, as a corollary, the untagged event sample, subsample
4, contains an excess of 3-jet events (Fig. 5a). Similar results
were obtained for the other jet algorithms (not shown).
  
Equations 1 were solved using 2- and 3-jet events defined in turn 
by each of the six 
jet algorithms to obtain the true 3-jet rates in \z0 \ra $q_l\overline{q_l}$,
\cc and \bb events, $R_3^{uds}$, $R_3^c$ and $R_3^b$ respectively.
Redefining $R_3^{all}$ = $\Sigma_{b,c,uds} f^i R_3^i$, the unfolded ratios 
$R_3^{uds}/R_3^{all}$, $R_3^c/R_3^{all}$ and $R_3^b/R_3^{all}$ 
are shown in Fig.~5b for 
comparison with the raw measured values shown in Fig.~5a. 

For the test of the flavor-independence of strong interactions it is more
convenient to consider the ratios of the 3-jet rates in heavy- and light-quark
events, namely $R_3^c/R_3^{uds}$ and $R_3^b/R_3^{uds}$.
These were derived from the unfolded $R_3^{uds}$, $R_3^c$ and $R_3^b$ values,
and the systematic errors on the ratios are considered in the next sections.

\vskip 1truecm

\noindent{\Large \bf 5. Experimental Systematic Errors}

\vskip .5truecm

\noindent
We considered sources of experimental systematic uncertainty that potentially 
affect our measurements of $R_3^c/R_3^{uds}$ and $R_3^b/R_3^{uds}$.
These may be divided into uncertainties in
modelling the detector and uncertainties on experimental measurements serving as
input parameters to the underlying physics modelling. In each case  
the error was evaluated by varying the appropriate parameter
in the Monte Carlo simulation, recalculating the matrices
$\varepsilon$, performing a new fit of Eq.~1 to the data, rederiving
values of $R_3^c/R_3^{uds}$ and $R_3^b/R_3^{uds}$, and taking the respective 
difference 
in results  relative to our standard procedure as the systematic uncertainty. 

In the category of detector modelling uncertainty we considered
the charged-particle tracking efficiency of the detector, as well as
the smearing applied to the simulated charged-particle impact
parameters in order to make the distributions agree with the data.
An extra tracking inefficiency of roughly 3.5\% was applied 
in the simulation in order to make the average number of charged tracks 
used for flavor-tagging agree with the data.
We repeated the analysis in turn without this efficiency correction, 
and with no impact-parameter smearing, in the simulation.

A large number of measured quantities relating to the production and decay
of charm and bottom hadrons are used as input to our simulation. 
In \bb events we have considered the uncertainties on: 
the average charged multiplicity of $B$-hadron decays, the $B$-hadron 
fragmentation function, the production rate of $b$-baryons, the $B$-meson and
$B$-baryon lifetimes, the inclusive production rate of $D^+$ mesons in
$B$-hadron decays, and 
the branching fraction for \z0 \ra \bb, $f^b$.
In \cc events we have considered the uncertainties on: 
the branching fraction $f^c$ for \z0 \ra \cc,  
the charmed hadron fragmentation function, 
the inclusive production rate of $D^+$ mesons, 
and the charged multiplicity of charmed hadron decays.
We also considered the rate of production of secondary \bb and \cc 
from gluon splitting in $q\bar{q}g$ events.
The values of these quantities used in our
simulation and the respective variations that we considered are listed in
Table~1. 

Statistical errors resulting from the finite size of the Monte Carlo event
sample were estimated by generating 1,000 toy Monte Carlo datasets of the
same size as that used in our data correction procedure, evaluating the
matrices $\varepsilon$ (Eq.~1) for each, unfolding the data, and calculating the
r.m.s. deviation of the distributions of the resulting 
$R_3^c/R_3^{uds}$ and $R_3^b/R_3^{uds}$ values.

As an example, for the E0 algorithm at $y_c$ = 0.02 the errors on 
$R_3^c/R_3^{uds}$ and $R_3^b/R_3^{uds}$
from the above sources are listed in Table~\ref{Table:syst}. 
The dominant physics contributions to $\delta R_3^b/R_3^{uds}$ result from 
limited knowledge 
of the average $B$-hadron decay multiplicity and the $B$-hadron fragmentation 
function.  The
uncertainties in $f^c$ and in the charmed hadron
fragmentation function produce the dominant variations
in $R_3^c/R_3^{uds}$. 
Contributions from $B$-hadron lifetimes, the fraction of 
$D^+$ in $B$ meson decays, $b$-baryon
production rates, and the charm hadron decay multiplicity are relatively small.

For each jet algorithm and $y_c$ value
all of the errors were added in quadrature to obtain a total experimental
systematic error on $R_3^c/R_3^{uds}$ and $R_3^b/R_3^{uds}$.
The choice of an optimal $y_c$ value is discussed in Section~6, and the 
combination of results from the six jet algorithms is discussed in Section~7.

\vskip 1truecm

\noindent{\Large \bf 6. Theoretical Uncertainties and Translation to \alp
Ratios}

\vskip .5truecm
 
\noindent
We considered sources of theoretical uncertainty that potentially affect 
our measurements. 
The ratios $R_3^c/R_3^{uds}$ and $R_3^b/R_3^{uds}$
derived in Section 4 were implicitly corrected for
the effects of hadronisation, and we have estimated the uncertainty in this
correction. Furthermore, the$\geq3$-jet rate in heavy-quark events is modified
relative to that in light-quark events by the effect of the
non-zero quark mass. This effect needs to be taken into account in the
translation between the jet-rate ratios and the corresponding ratios of
strong couplings $\alpha_s^c/\alpha_s^{uds}$ and $\alpha_s^b/\alpha_s^{uds}$.
We have used \oalpsq calculations
to perform the mass-dependent translation, and have estimated 
the related uncertainties due to the value of the $b$-quark mass, as well as
higher-order perturbative QCD contributions.

\vskip 1truecm

\noindent{\large \bf 6.1 Hadronisation Uncertainties}

\vskip .5truecm

\noindent
The intrinsically non-perturbative process by which quarks and gluons fragment
into the observed final-state hadrons cannot currently be calculated in QCD.
Phenomenological models of hadronisation have been developed over the past
few decades and have been implemented in Monte Carlo event-generator programs
to facilitate comparison with experimental data. We have used the models
implemented in the JETSET 7.4 and HERWIG 5.9~\cite{herwig} programs to study
hadronisation effects; these models have been extensively studied and tuned to
provide a good description of detailed properties of hadronic final states in
\epa; for a review of studies at the \z0 resonance see~\cite{knowles}.
Our standard simulation based on JETSET 7.4 was used to evaluate the 
efficiency and purity of the event-flavor tagging, 
as described in Section 4, as well as for the study of experimental
systematic errors described in Section 5. 

We investigated hadronisation uncertainties by calculating from the
Monte Carlo-generated event sample the ratios:
$$
r_i \quad \equiv \quad \left({R_3^i \over {R_3^{uds}}}\right)_{parton}/
\left({R_3^i \over {R_3^{uds}}}\right)_{hadron}
$$
where $i$ = $c$ or $b$, $parton$ refers to the calculation of the quantity in 
brackets at the parton-level, and $hadron$ refers to the corresponding 
hadron-level calculation using stable final-state particles. 
We recalculated these ratios by changing in turn the parameters $Q_0$ and
$\sigma_q$ in the JETSET program\footnote{$Q_0$ (GeV) controls the minimum 
virtual
mass allowed for partons in the parton shower; we considered a variation around
the central value, 1.0, of $^{+1.0}_{-0.5}$. $\sigma_q$
(GeV/$c$) is the width of the Gaussian distribution used to assign
transverse momentum, w.r.t. the color field, to quarks and antiquarks 
produced in the fragmentation process; we considered a variation around
the central value, 0.39, of $\pm0.04$.}
and generating 1-million-event samples. 
We also recalculated these ratios by using the HERWIG 5.9 program
with default parameter settings. For each variation we evaluated the fractional
deviation $\Delta r_i$ w.r.t. the standard value:
$$
\Delta r_i\quad = \quad {(r_i^{\prime} - r_i)\over {r_i}},
$$
and the corresponding deviations on $R_3^i/R_3^{uds}$.
As an example, for the E0 algorithm and $y_c$ = 0.02 the deviations are listed
in Table~1. The deviations were added in quadrature to
define the systematic error on $R_3^i / R_3^{uds}$
due to hadronisation uncertainties.

\vskip 1truecm
 
\noindent{\large \bf 6.2 Choice of $y_c$ Values}

\vskip .5truecm

\noindent
For each jet algorithm and $y_c$ value the statistical and experimental 
systematic errors
and hadronisation uncertainty on each $R_3^i / R_3^{uds}$ were added in 
quadrature. No strong dependence of this combined 
error on $y_c$ was observed~\cite{thesis}, but an `optimal' $y_c$
value for each algorithm was then identified that corresponded with 
the smallest error.
In the case of the E and G algorithms slightly larger $y_c$ values were chosen
so as to ensure that the \oalpsq calculations for massive quarks were
reliable~\cite{Brandenburg}.
The chosen $y_c$ value for each algorithm is listed in Table~2, together with
the corresponding values of the ratios $R_3^c / R_3^{uds}$ and 
$R_3^b / R_3^{uds}$, as well as the statistical and
experimental-systematic errors and hadronisation uncertainties.

\vskip 1truecm
 
\noindent{\large \bf 6.3 Translation to \alp Ratios}

\vskip .5truecm

\noindent
The test of the flavor-independence of strong
interactions can be expressed in terms of the ratios
$\alpha_s^i/\alpha_s^{uds}$ ($i$ = $c$ or $b$).
Recalling that with our definition $R_3$ is the rate of production of 3 or more
jets, $\alpha_s^i/\alpha_s^{uds}$
can be derived from the respective measured ratio $R_3^i / R_3^{uds}$ using the 
next-to-leading-order perturbative QCD calculation:
\begin{eqnarray}  
{R_3^i\over {R_3^{uds}}}\quad = \quad 
{{A^i\,\alpha_s^i\ + \bigl[B^i+C^i\bigr]\; (\alpha_s^i)^2 \;+\;
O(({\alpha_s^i})^3)}
\over {A^{uds}\,\alpha_s^{uds}\ + \bigl[B^{uds}+C^{uds}\bigr]\; 
(\alpha_s^{uds})^2\;+\; O(({\alpha_s^{uds}})^3)}}
\end{eqnarray}  
where the coefficients $A$, $B$ and $C$ represent, respectively,
the leading-order (LO) perturbative QCD coefficient for the 3-jet rate, 
the next-to-leading-order (NLO) coefficient for this rate, and the 
leading-order coefficient 
for the 4-jet rate. Next-to-leading-order contributions to the 4-jet rate, and
contributions from $\geq5$-jet rates, are represented by the terms of
O$(\alpha_s^3)$. 
These coefficients depend implicitly upon the jet algorithm as well as on the
scaled-invariant-mass-squared jet resolution parameter $y_c$; for clarity these
dependences have been omitted from the notation.
For massless quarks calculations of the coefficients $A$, $B$ and $C$
have been available for many years~\cite{Siggi,KN}.

For many observables
at the $Z^0$ pole the quark mass appears in terms proportional to 
the ratio  $m_q^2/M_Z^2$, and
  the effects of non-zero quark mass can be neglected.
   For the jet rates, however, 
  mass effects can enter via terms proportional to 
$m_q^2/(y_c M_Z^2)$. For $b$-quarks these terms can contribute at the
 O$(5\%)$ level for typical values of $y_c$
used in jet clustering. Therefore,
  the $\geq3$-jet rate in heavy-quark events is expected to be 
 modified relative to that in light-quark events both by the diminished 
 phase-space for gluon emission due to the quark mass, as well as by kinematic
effects in the definition of the jet clustering schemes.
Such mass effects for jet rates have very recently been 
calculated~\cite{Aachen,Rodrigo} at NLO in perturbative QCD\footnote{In our
previous study~\cite{MHalphas} only the relevant tree-level calculations
for 3-jet and 4-jet final-states were available.}, and the quark-mass dependence
can be expressed in terms of the {\it running} mass $m_b(M_{Z^0})$.
The Aachen group has evaluated~\cite{Brandenburg} the terms $A^b$, $B^b$
and $C^b$ for massive $b$-quarks at our preferred values of $y_c$;
these are listed in Table~3. 

For illustration, the measured ratios $R_3^c / R_3^{uds}$ and 
$R_3^b / R_3^{uds}$, are shown in Fig.~\ref{Fig:results}(a). 
$R_3^b / R_3^{uds}$ lies above unity for the 
E, E0, P and P0 algorithms, and below unity for the D and G algorithms; note
that all six data points are highly correlated with each other, so that the
differences between algorithms are more significant than naively implied by 
the statistical errors displayed. For comparison, the corresponding QCD
calculations of $R_3^b / R_3^{uds}$ 
are also shown in Fig.~\ref{Fig:results}(a), under the {\it assumption} of a 
flavor-independent strong coupling with an input value of \alpmzsq = 0.118,
for $m_b(M_{Z^0})$ = $3.0\pm0.5$ GeV/$c^2$.
Under this assumption the calculations are in
good agreement with the data, and the data clearly demonstrate the effects of
the non-zero $b$-quark mass, which are larger than the statistical error. 
For the translation from $R_3^b / R_3^{uds}$ to $\alpha_s^b/\alpha_s^{uds}$
we used a value of the running $b$-quark mass $m_b(M_{Z^0}) = 3.0$ GeV$/c^2$. 

For $c$-quarks mass effects are expected to be O$(1\%)$ or 
less~\cite{Brandenburg}, which is much smaller than our statistical error of 
roughly 4\% on $R_3^c/R_3^{uds}$. The effects of non-zero $c$-quark mass, and 
of the light-quark masses, will hence be neglected here.
We used values of $A^{uds}$, $B^{uds}$ and $C^{uds}$ from Ref.~\cite{Siggi}.
                     
Eqns.~(2) were solved to obtain the ratios $\alpha_s^c/\alpha_s^{uds}$
and $\alpha_s^b/\alpha_s^{uds}$ for each jet algorithm. These ratios are listed
in Table~2, together with the corresponding statistical and experimental
systematic errors, and the hadronisation uncertainties.
We then evaluated sources of uncertainty in this translation procedure.
From an operational point of view these affect the
values of the coefficients $A$, $B$ and $C$ used for the translation.
For each variation considered the relevant $A$, $B$ or $C$ were 
reevaluated, the ratios  $\alpha_s^i/\alpha_s^{uds}$ 
were rederived, and the deviation w.r.t. the central value 
was assigned as a systematic uncertainty.

We considered a variation of $\pm0.5$ GeV$/c^2$ about the central value of 
the running $b$-quark mass $m_b(M_{Z^0}) = 3.0 $GeV$/c^2$. 
This corresponds to the range $3.62<m_b(m_b)<5.06$ GeV$/c^2$ and covers
generously the 
values~\cite{Rodrigo} determined from the $\Upsilon$ system using QCD sum 
rules, $4.13\pm0.06$ GeV$/c^2$, as well as using lattice QCD, $4.15\pm0.20$ 
GeV$/c^2$.
It is also consistent with the recent DELPHI measurement of the running mass: 
 $m_b(M_{Z^0}) = 2.67\pm0.25(stat.)\pm0.34(frag.)\pm0.27(theo.)$ 
GeV$/c^2$~\cite{DELPHI97}.
The numerical accuracy on the coefficients $A$, $B$, and $C$ is
 in all cases negligibly small on the scale of the experimental statistical 
 errors.

We considered the effects of the uncalculated higher-order terms in Eq.~(2).
In these ratios the
effects of such higher-order contributions will tend to cancel. Nevertheless we
have attempted to evaluate the residual uncertainty due to 
these contributions. We first considered 3-jet contributions and varied the
NLO coefficient $B$; for each jet algorithm we varied simultaneously the 
renormalisation scale $\mu$ and $\alpha_s^{uds}$ in the ranges allowed 
by fits to the flavor-inclusive differential 2-jet 
rate~\cite{sldalphas}\footnote{
Heavy-quark mass and possible flavor-dependent effects
are negligible on the scale of the large errors considered on $\alpha_s^{uds}$
for this purpose.}.
In addition, we considered next-to-leading order (NLO)
 contributions to the 4-jet rate. Although these enter
formally at \oalpc in Eq.~(2),  operationally they may
be estimated by variation of the LO coefficient $C^i$. Since the 4-jet rate
has been calculated recently complete at NLO for massless quarks~\cite{lance},
these terms can be estimated reliably. For our jet algorithms and $y_c$
values Dixon has evaluated the LO and NLO 4-jet contributions~\cite{lancepriv}.
Based on these calculations we varied the coefficient $C$ by $\pm$100\%.
For each jet algorithm, at the chosen $y_c$ value, 
the measured contribution to $R_3$ from 
 $\geq$5-jet states was smaller than 1\% 
 and the corresponding \oalpc contributions to
  Eq.~(2)  were neglected.

These uncertainties are summarised in Table~\ref{Table:translation}.
The deviations for each variation considered were 
added in quadrature to define a total translation uncertainty on 
$\alpha_s^c/\alpha_s^{uds}$ and $\alpha_s^b/\alpha_s^{uds}$, listed in 
Table~2.
 
\vskip 1truecm
 
\noindent{\Large \bf 7. Comparison of \alp Ratios}

\vskip .5truecm

\noindent
The $\alpha_s^c/\alpha_s^{uds}$ and $\alpha_s^b/\alpha_s^{uds}$ 
ratios are summarised in Fig.~\ref{Fig:results}b. 
It can be seen that the 
ratios determined using the different jet algorithms are in good agreement
with one another.

For each jet algorithm $n$, the 
statistical and experimental systematic errors were added in quadrature with
the hadronisation and translation uncertainties 
(Table~2) to define a total error $\sigma_n^i$ on $\alpha_s^i/\alpha_s^{uds}$
($i$ = $c$ or $b$). For each flavor a single value of 
$\alpha_s^i/\alpha_s^{uds}$ was then defined by taking the weighted
average of the results over the six jet algorithms: 
\begin{eqnarray}
\alpha_s^i/\alpha_s^{uds}  = \sum_n w_n^i (\alpha_s^i/\alpha_s^{uds})_n
\end{eqnarray}
 where $w_n^i$ is the weight for each algorithm:
\begin{eqnarray}
 w_n^i &=& \frac{1/{\sigma_n^i}^2} {\sum_n 1/{\sigma_n^i}^2}
\end{eqnarray}
The average statistical and experimental systematic errors were each
computed from:
\begin{eqnarray}
 \overline {\sigma^i} = \sqrt { \sum_{nm} E^i_{nm} w^i_n w^i_m }
\end{eqnarray}
 where $E^i$ is the $6\times 6$ covariant matrix with elements:
 \begin{eqnarray}
  E^i_{nm}= \sigma^i_n \sigma^i_m
 \end{eqnarray}
and 100\% correlation was conservatively assumed among algorithms.
The average translation and hadronisation uncertainties were calculated in a
similar fashion. 
We then calculated the r.m.s. deviation on 
$\alpha_s^c/\alpha_s^{uds}$ and $\alpha_s^b/\alpha_s^{uds}$, 
shown in Table~2, and
assigned this scatter between the results from different algorithms
as an additional theoretical uncertainty.
The average translation and hadronisation uncertainties were added
in quadrature together with the
r.m.s. deviation to define the total theoretical uncertainty.

We obtained:
 \begin{eqnarray*}
  \alpha_s^{c}/\alpha_s^{uds}   &=& 1.036\pm0.043     (stat.) 
                                    ^{+0.041}_{-0.045}(syst.) 
                                    ^{+0.020}_{-0.018}(theory) \\ 
  \alpha_s^{b}/\alpha_s^{uds}   &=& 1.004\pm0.018(stat.) 
                                    ^{+0.026}_{-0.031}(syst.) 
                                    ^{+0.018}_{-0.029}(theory). \\ 
 \end{eqnarray*}
The theoretical uncertainties are only slightly smaller than the respective
experimental systematic errors, and comprise roughly equal contributions from
the hadronisation and translation uncertainties, as well as from the r.m.s. 
deviation over the six jet algorithms.

\vskip 1truecm
 
\noindent{\Large \bf 8. Cross-checks}

\vskip .5truecm

\noindent
We performed a number of cross-checks on these results. 
First, we varied the event selection requirements.
The thrust-axis containment cut was varied in the range 
$0.65<|\cos\theta_T| < 0.75$, the minimum number of charged tracks
required was increased from 7 to 8, and the total charged-track energy
requirement was increased from 20 to 22 GeV.
In each case 
results consistent with the standard selection were obtained. 

Next, we included in the unfolding procedure (Eq.~(1) and Section 4)
 the `untagged' event sample, subsample 4 (Section 3),
 whose flavor composition is similar to the natural composition in 
flavor-inclusive \z0 decay events, and repeated the analysis to derive new
values of $\alpha_s^{c}/\alpha_s^{uds}$ and $\alpha_s^{b}/\alpha_s^{uds}$. 
In addition, we repeated the unfolding and, instead of fixing them to 
Standard Model values (Table~1), allowed the \z0 \ra \cc and
\z0 \ra \bb branching fractions to float in the fit of Eq.~(1).
In both cases results consistent with the standard procedure were 
obtained~\cite{thesis}. 

We also considered variations of the flavor-tagging scheme based on 
reconstructed secondary vertices. In each case we repeated the 
analysis described in Sections 4-7 and derived new values of 
$\alpha_s^{c}/\alpha_s^{uds}$ and $\alpha_s^{b}/\alpha_s^{uds}$.
Firstly, 
we used more efficient tags for primary \bb and \cc events.
We applied the scheme described in
Section 3, but with a looser definition of region (A) to include vertices with 
$M_{vtx}>1.8$ or $P_{vtx}+10<15M_{vtx}$.
We also removed the cut on the vertex impact parameter,
$\delta_{vtx}$, used to define the $c$-tagged sample, and region (B) was
redefined
to comprise only events with $N_{sig}$ $\geq1$ and containing a SV with 
$P_{vtx}>5$ $\oplus$ $P_{vtx}+10>15M_{vtx}$.
Second, we repeated this modified scheme, but increased the efficiency 
for light-quark tagging by requiring tracks that miss the IP by at least
3$\sigma_d$ to be counted in $N_{sig}$ for the definition of
the $uds$-tagged sample. 
Third, we did not use vertex momentum information for the tag definitions; we
used instead only vertex mass information to define region (A):
$M_{vtx}>1.8$, and Region (B): $M_{vtx}<1.8$, with the $uds$-tagged 
sample defined as in Section~4. Finally, we tried a variation
in which we used event hemispheres as a basis for flavor-tagging, rather than
jets as defined in Section 3; this tag is similar to that used in our recent
study of the branching fraction for \z0 \ra \bb~\cite{sldnewrb}.
In all cases results statistically consistent with our standard analysis were
obtained~\cite{thesis}.

We also performed an analysis using a similar flavor-tagging technique to that
reported in our previous publication~\cite{MHalphas}.
We counted the number of tracks per event, $N_{sig}$, 
that miss the IP by $d>3\sigma_d$. 
This distribution is shown in Fig.~7; 
the data are well described by
our Monte Carlo  simulation. For the simulation,  the contributions of events 
of different quark flavors are shown separately. 
The leftmost bin contains predominantly events
containing primary $u$, $d$, or $s$ quarks, while the rightmost bins 
contain a pure sample of events containing primary $b$ quarks. 
the event sample was divided accordingly into five subsamples according to the
number of `significant' tracks: 
(i) $N_{sig}=0$, (ii) $N_{sig}=1$, (iii) $N_{sig}=2$, (iv) $N_{sig}=3$, and (v) 
$N_{sig}\geq 4$. 
A similar formalism to that defined by Eq.~(1) was applied using
$5\times3$ matrices $\varepsilon$ and yielded values 
of \ruds, \rc and \rb consistent with those obtained in Sections 4 and 5, 
but with larger statistical and systematic errors.     Furthermore,
we also applied a simpler version of this technique in which subsamples
(ii), (iii) and (iv) were combined into a single $c$-tagged sample and a 
$3\times3$
flavor unfolding was performed. Again, this yielded values 
of \ruds, \rc and \rb consistent with those obtained in Sections 4 and 5, 
but with larger statistical and systematic errors~\cite{thesis}.     

\vskip 1truecm

\noindent{\Large \bf 9. Summary and Discussion} 

\vskip .5truecm
 
\noindent
We have used hadron lifetime and mass information 
to separate hadronic $Z^0$ decays into tagged \bb, \cc and light-quark event
samples with high efficiency and purity,  and small
 bias against events containing hard gluon radiation.
From a comparison of the rates of multijet events in these
samples, we obtained:
 \begin{eqnarray*}
  \alpha_s^{c}/\alpha_s^{uds}   &=& 1.036\pm0.043     (stat.) 
                                    ^{+0.041}_{-0.045}(syst.) 
                                    ^{+0.020}_{-0.018}(theory) \\ 
  \alpha_s^{b}/\alpha_s^{uds}   &=& 1.004\pm0.018(stat.) 
                                    ^{+0.026}_{-0.031}(syst.) 
                                    ^{+0.018}_{-0.029}(theory). \\ 
 \end{eqnarray*}
We find that the strong coupling is independent of
quark flavor within our sensitivity. 

For comparison with our previous result and with other experiments 
one can discuss the test of flavor-independence in terms of the ratios 
$\alpha_s^{uds}/\alpha_s^{all}$, $\alpha_s^{c}/\alpha_s^{all}$ and
$\alpha_s^{b}/\alpha_s^{all}$, although these quantities, by construction, are
not independent of each another. We performed a similar analysis to that 
described in Sections~6
and 7 using, instead of $R_3^{c}/R_3^{uds}$ and $R_3^{b}/R_3^{uds}$,
our measured values of $R_3^{uds}/R_3^{all}$, 
$R_3^{c}/R_3^{all}$ and $R_3^{b}/R_3^{all}$ (Section~4) as a starting point.
We obtained:
\begin{eqnarray*}
\alpha_s^{uds}/\alpha_s^{all} &=& 0.987\pm0.010    (stat.) 
                                 ^{+0.012}_{-0.010}(syst.) 
                                  ^{+0.009}_{-0.008}(theory)  \\ 
\alpha_s^{c}/\alpha_s^{all}   &=& 1.023\pm0.034    (stat.) 
                                 ^{+0.032}_{-0.036}(syst.) 
                                 ^{+0.018}_{-0.014}(theory)  \\ 
\alpha_s^{b}/\alpha_s^{all}   &=& 0.993\pm0.016    (stat.) 
                                 ^{+0.020}_{-0.023}(syst.) 
                                 ^{+0.019}_{-0.027}(theory).  \\ 
\end{eqnarray*}
These results are consistent with, and supersede, our previous 
measurements~\cite{MHalphas},
and are substantially more precise; they are also consistent with
measurements performed at LEP using different flavor-tagging 
techniques~\cite{DELPHI97,LEPalphas,aleph,OPAL}. A summary of these results is 
given in Fig.~8. Our comprehensive study, involving six jet-finding algorithms, 
and the inclusion of the 
resulting r.m.s. deviations of results as additional uncertainties, represents a
conservative procedure.
 
\section*{Acknowledgements}
We thank the personnel of the SLAC accelerator department and the technical
staffs of our collaborating institutions for their outstanding efforts
on our behalf. 
We also thank A.~Brandenburg, P.~Uwer and L.~Dixon 
for performing onerous QCD calculations for this analysis and for helpful
contributions, as well as T.~Rizzo for many useful discussions. 
 
\vskip 1truecm
  
\section*{$^{**}$List of Authors}


%
%
%
\begin{center}
\def\iADEL{$^{(1)}$}
\def\iAOMORI{$^{(2)}$}
\def\iBOLO{$^{(3)}$}
\def\iBRUN{$^{(4)}$}
\def\iBU{$^{(5)}$}
\def\iCINC{$^{(6)}$}
\def\iCOLO{$^{(7)}$}
\def\iCOLU{$^{(8)}$}
\def\iCSU{$^{(9)}$}
\def\iFERR{$^{(10)}$}
\def\iFRAS{$^{(11)}$}
\def\iILLI{$^{(12)}$}
\def\iLBL{$^{(13)}$}
\def\iMASS{$^{(14)}$}
\def\iMISSI{$^{(15)}$}
\def\iMIT{$^{(16)}$}
\def\iNAGO{$^{(17)}$}
\def\iOREG{$^{(18)}$}
\def\iOXF{$^{(19)}$}
\def\iPADO{$^{(20)}$}
\def\iPERU{$^{(21)}$}
\def\iPISA{$^{(22)}$}
\def\iRAL{$^{(23)}$}
\def\iRUTG{$^{(24)}$}
\def\iSLAC{$^{(25)}$}
\def\iSOGA{$^{(26)}$}
\def\iSOONG{$^{(27)}$}
\def\iTENN{$^{(28)}$}
\def\iTOHO{$^{(29)}$}
\def\iUCSB{$^{(30)}$}
\def\iUCSC{$^{(31)}$}
\def\iVAND{$^{(32)}$}
\def\iWASH{$^{(33)}$}
\def\iWISC{$^{(34)}$}
\def\iYALE{$^{(35)}$}

  \baselineskip=.75\baselineskip  
\mbox{K. Abe\unskip,\iTOHO}
\mbox{K.  Abe\unskip,\iNAGO}
\mbox{T. Abe\unskip,\iSLAC}
\mbox{T.  Akagi\unskip,\iSLAC}
\mbox{N. J. Allen\unskip,\iBRUN}
\mbox{A. Arodzero\unskip,\iOREG}
\mbox{D. Aston\unskip,\iSLAC}
\mbox{K.G. Baird\unskip,\iMASS}
\mbox{C. Baltay\unskip,\iYALE}
\mbox{H.R. Band\unskip,\iWISC}
\mbox{T.L. Barklow\unskip,\iSLAC}
\mbox{J.M. Bauer\unskip,\iMISSI}
\mbox{A.O. Bazarko\unskip,\iCOLU}
\mbox{G. Bellodi\unskip,\iOXF}
\mbox{A.C. Benvenuti\unskip,\iBOLO}
\mbox{G.M. Bilei\unskip,\iPERU}
\mbox{D. Bisello\unskip,\iPADO}
\mbox{G. Blaylock\unskip,\iMASS}
\mbox{J.R. Bogart\unskip,\iSLAC}
\mbox{T. Bolton\unskip,\iCOLU}
\mbox{G.R. Bower\unskip,\iSLAC}
\mbox{J. E. Brau\unskip,\iOREG}
\mbox{M. Breidenbach\unskip,\iSLAC}
\mbox{W.M. Bugg\unskip,\iTENN}
\mbox{D. Burke\unskip,\iSLAC}
\mbox{T.H. Burnett\unskip,\iWASH}
\mbox{P.N. Burrows\unskip,\iOXF}
\mbox{A. Calcaterra\unskip,\iFRAS}
\mbox{D.O. Caldwell\unskip,\iUCSB}
\mbox{D. Calloway\unskip,\iSLAC}
\mbox{B. Camanzi\unskip,\iFERR}
\mbox{M. Carpinelli\unskip,\iPISA}
\mbox{R. Cassell\unskip,\iSLAC}
\mbox{R. Castaldi\unskip,\iPISA}
\mbox{A. Castro\unskip,\iPADO}
\mbox{M. Cavalli-Sforza\unskip,\iUCSC}
\mbox{A. Chou\unskip,\iSLAC}
\mbox{H.O. Cohn\unskip,\iTENN}
\mbox{J.A. Coller\unskip,\iBU}
\mbox{M.R. Convery\unskip,\iSLAC}
\mbox{V. Cook\unskip,\iWASH}
\mbox{R.F. Cowan\unskip,\iMIT}
\mbox{D.G. Coyne\unskip,\iUCSC}
\mbox{G. Crawford\unskip,\iSLAC}
\mbox{C.J.S. Damerell\unskip,\iRAL}
\mbox{M. Daoudi\unskip,\iSLAC}
\mbox{N. de Groot\unskip,\iSLAC}
\mbox{R. Dell'Orso\unskip,\iPERU}
\mbox{P.J. Dervan\unskip,\iBRUN}
\mbox{R. de Sangro\unskip,\iFRAS}
\mbox{M. Dima\unskip,\iCSU}
\mbox{A. D'Oliveira\unskip,\iCINC}
\mbox{D.N. Dong\unskip,\iMIT}
\mbox{R. Dubois\unskip,\iSLAC}
\mbox{B.I. Eisenstein\unskip,\iILLI}
\mbox{V. Eschenburg\unskip,\iMISSI}
\mbox{E. Etzion\unskip,\iWISC}
\mbox{S. Fahey\unskip,\iCOLO}
\mbox{D. Falciai\unskip,\iFRAS}
\mbox{J.P. Fernandez\unskip,\iUCSC}
\mbox{M.J. Fero\unskip,\iMIT}
\mbox{R. Frey\unskip,\iOREG}
\mbox{G. Gladding\unskip,\iILLI}
\mbox{E.L. Hart\unskip,\iTENN}
\mbox{J.L. Harton\unskip,\iCSU}
\mbox{A. Hasan\unskip,\iBRUN}
\mbox{K. Hasuko\unskip,\iTOHO}
\mbox{S. J. Hedges\unskip,\iBU}
\mbox{S.S. Hertzbach\unskip,\iMASS}
\mbox{M.D. Hildreth\unskip,\iSLAC}
\mbox{M.E. Huffer\unskip,\iSLAC}
\mbox{E.W. Hughes\unskip,\iSLAC}
\mbox{X.Huynh\unskip,\iSLAC}
\mbox{M. Iwasaki\unskip,\iOREG}
\mbox{D. J. Jackson\unskip,\iRAL}
\mbox{P. Jacques\unskip,\iRUTG}
\mbox{J.A. Jaros\unskip,\iSLAC}
\mbox{Z.Y. Jiang\unskip,\iSLAC}
\mbox{A.S. Johnson\unskip,\iSLAC}
\mbox{J.R. Johnson\unskip,\iWISC}
\mbox{R.A. Johnson\unskip,\iCINC}
\mbox{R. Kajikawa\unskip,\iNAGO}
\mbox{M. Kalelkar\unskip,\iRUTG}
\mbox{Y. Kamyshkov\unskip,\iTENN}
\mbox{H.J. Kang\unskip,\iRUTG}
\mbox{I. Karliner\unskip,\iILLI}
\mbox{Y. D. Kim\unskip,\iSOGA}
\mbox{M.E. King\unskip,\iSLAC}
\mbox{R.R. Kofler\unskip,\iMASS}
\mbox{R.S. Kroeger\unskip,\iMISSI}
\mbox{M. Langston\unskip,\iOREG}
\mbox{D.W.G. Leith\unskip,\iSLAC}
\mbox{V. Lia\unskip,\iMIT}
\mbox{X. Liu\unskip,\iUCSC}
\mbox{M.X. Liu\unskip,\iYALE}
\mbox{M. Loreti\unskip,\iPADO}
\mbox{H.L. Lynch\unskip,\iSLAC}
\mbox{G. Mancinelli\unskip,\iRUTG}
\mbox{S. Manly\unskip,\iYALE}
\mbox{G. Mantovani\unskip,\iPERU}
\mbox{T.W. Markiewicz\unskip,\iSLAC}
\mbox{T. Maruyama\unskip,\iSLAC}
\mbox{H. Masuda\unskip,\iSLAC}
\mbox{A.K. McKemey\unskip,\iBRUN}
\mbox{B.T. Meadows\unskip,\iCINC}
\mbox{G. Menegatti\unskip,\iFERR}
\mbox{R. Messner\unskip,\iSLAC}
\mbox{P.M. Mockett\unskip,\iWASH}
\mbox{K.C. Moffeit\unskip,\iSLAC}
\mbox{T.B. Moore\unskip,\iYALE}
\mbox{M.Morii\unskip,\iSLAC}
\mbox{D. Muller\unskip,\iSLAC}
\mbox{T. Nagamine\unskip,\iTOHO}
\mbox{S. Narita\unskip,\iTOHO}
\mbox{U. Nauenberg\unskip,\iCOLO}
\mbox{M. Nussbaum\unskip,\iCINC}
\mbox{N.Oishi\unskip,\iNAGO}
\mbox{D. Onoprienko\unskip,\iTENN}
\mbox{L.S. Osborne\unskip,\iMIT}
\mbox{R.S. Panvini\unskip,\iVAND}
\mbox{C. H. Park\unskip,\iSOONG}
\mbox{T.J. Pavel\unskip,\iSLAC}
\mbox{I. Peruzzi\unskip,\iFRAS}
\mbox{M. Piccolo\unskip,\iFRAS}
\mbox{L. Piemontese\unskip,\iFERR}
\mbox{E. Pieroni\unskip,\iPISA}
\mbox{R.J. Plano\unskip,\iRUTG}
\mbox{R. Prepost\unskip,\iWISC}
\mbox{C.Y. Prescott\unskip,\iSLAC}
\mbox{G.D. Punkar\unskip,\iSLAC}
\mbox{J. Quigley\unskip,\iMIT}
\mbox{B.N. Ratcliff\unskip,\iSLAC}
\mbox{J. Reidy\unskip,\iMISSI}
\mbox{P.L. Reinertsen\unskip,\iUCSC}
\mbox{L.S. Rochester\unskip,\iSLAC}
\mbox{P.C. Rowson\unskip,\iSLAC}
\mbox{J.J. Russell\unskip,\iSLAC}
\mbox{O.H. Saxton\unskip,\iSLAC}
\mbox{T. Schalk\unskip,\iUCSC}
\mbox{R.H. Schindler\unskip,\iSLAC}
\mbox{B.A. Schumm\unskip,\iUCSC}
\mbox{J. Schwiening\unskip,\iSLAC}
\mbox{S. Sen\unskip,\iYALE}
\mbox{V.V. Serbo\unskip,\iWISC}
\mbox{M.H. Shaevitz\unskip,\iCOLU}
\mbox{J.T. Shank\unskip,\iBU}
\mbox{G. Shapiro\unskip,\iLBL}
\mbox{D.J. Sherden\unskip,\iSLAC}
\mbox{K. D. Shmakov\unskip,\iTENN}
\mbox{N.B. Sinev\unskip,\iOREG}
\mbox{S.R. Smith\unskip,\iSLAC}
\mbox{M. B. Smy\unskip,\iCSU}
\mbox{J.A. Snyder\unskip,\iYALE}
\mbox{H. Staengle\unskip,\iCSU}
\mbox{A. Stahl\unskip,\iSLAC}
\mbox{P. Stamer\unskip,\iRUTG}
\mbox{R. Steiner\unskip,\iADEL}
\mbox{H. Steiner\unskip,\iLBL}
\mbox{D. Su\unskip,\iSLAC}
\mbox{F. Suekane\unskip,\iTOHO}
\mbox{A. Sugiyama\unskip,\iNAGO}
\mbox{S. Suzuki\unskip,\iNAGO}
\mbox{M. Swartz\unskip,\iSLAC}
\mbox{F.E. Taylor\unskip,\iMIT}
\mbox{J. Thom\unskip,\iSLAC}
\mbox{E. Torrence\unskip,\iMIT}
\mbox{N. K. Toumbas\unskip,\iSLAC}
\mbox{A.I. Trandafir\unskip,\iMASS}
\mbox{J.D. Turk\unskip,\iYALE}
\mbox{T. Usher\unskip,\iSLAC}
\mbox{C. Vannini\unskip,\iPISA}
\mbox{J. Va'vra\unskip,\iSLAC}
\mbox{E. Vella\unskip,\iSLAC}
\mbox{J.P. Venuti\unskip,\iVAND}
\mbox{R. Verdier\unskip,\iMIT}
\mbox{P.G. Verdini\unskip,\iPISA}
\mbox{S.R. Wagner\unskip,\iSLAC}
\mbox{D. L. Wagner\unskip,\iCOLO}
\mbox{A.P. Waite\unskip,\iSLAC}
\mbox{C. Ward\unskip,\iBRUN}
\mbox{S.J. Watts\unskip,\iBRUN}
\mbox{A.W. Weidemann\unskip,\iTENN}
\mbox{E. R. Weiss\unskip,\iWASH}
\mbox{J.S. Whitaker\unskip,\iBU}
\mbox{S.L. White\unskip,\iTENN}
\mbox{F.J. Wickens\unskip,\iRAL}
\mbox{D.C. Williams\unskip,\iMIT}
\mbox{S.H. Williams\unskip,\iSLAC}
\mbox{S. Willocq\unskip,\iSLAC}
\mbox{R.J. Wilson\unskip,\iCSU}
\mbox{W.J. Wisniewski\unskip,\iSLAC}
\mbox{J. L. Wittlin\unskip,\iMASS}
\mbox{M. Woods\unskip,\iSLAC}
\mbox{T.R. Wright\unskip,\iWISC}
\mbox{J. Wyss\unskip,\iPADO}
\mbox{R.K. Yamamoto\unskip,\iMIT}
\mbox{X. Yang\unskip,\iOREG}
\mbox{J. Yashima\unskip,\iTOHO}
\mbox{S.J. Yellin\unskip,\iUCSB}
\mbox{C.C. Young\unskip,\iSLAC}
\mbox{H. Yuta\unskip,\iAOMORI}
\mbox{G. Zapalac\unskip,\iWISC}
\mbox{R.W. Zdarko\unskip,\iSLAC}
\mbox{J. Zhou\unskip.\iOREG}

\it
  \vskip \baselineskip                   
  \vskip \baselineskip        
  \baselineskip=.75\baselineskip   
\iADEL
  Adelphi University,
  South Avenue-   Garden City,NY 11530, \break
\iAOMORI
  Aomori University,
  2-3-1 Kohata, Aomori City, 030 Japan, \break
\iBOLO
  INFN Sezione di Bologna,
  Via Irnerio 46    I-40126 Bologna  (Italy), \break
\iBRUN
  Brunel University,
  Uxbridge, Middlesex - UB8 3PH United Kingdom, \break
\iBU
  Boston University,
  590 Commonwealth Ave. - Boston,MA 02215, \break
\iCINC
  University of Cincinnati,
  Cincinnati,OH 45221, \break
\iCOLO
  University of Colorado,
  Campus Box 390 - Boulder,CO 80309, \break
\iCOLU
  Columbia University,
  Nevis Laboratories  P.O.Box 137 - Irvington,NY 10533, \break
\iCSU
  Colorado State University,
  Ft. Collins,CO 80523, \break
\iFERR
  INFN Sezione di Ferrara,
  Via Paradiso,12 - I-44100 Ferrara (Italy), \break
\iFRAS
  Lab. Nazionali di Frascati,
  Casella Postale 13   I-00044 Frascati (Italy), \break
\iILLI
  University of Illinois,
  1110 West Green St.  Urbana,IL 61801, \break
\iLBL
  Lawrence Berkeley Laboratory,
  Dept.of Physics 50B-5211 University of California-  Berkeley,CA 94720, \break
\iMASS
  University of Massachusetts,
  Amherst,MA 01003, \break
\iMISSI
  University of Mississippi,
  University,MS 38677, \break
\iMIT
  Massachusetts Institute of Technology,
  77 Massachussetts Avenue  Cambridge,MA 02139, \break
\iNAGO
  Nagoya University,
  Nagoya 464 Japan, \break
\iOREG
  University of Oregon,
  Department of Physics  Eugene,OR 97403, \break
\iOXF
  Oxford University,
  Oxford, OX1 3RH, United Kingdom, \break
\iPADO
  Universita di Padova,
  Via F. Marzolo,8   I-35100 Padova (Italy), \break
\iPERU
  Universita di Perugia, Sezione INFN,
  Via A. Pascoli  I-06100 Perugia (Italy), \break
\iPISA
  INFN, Sezione di Pisa,
  Via Livornese,582/AS  Piero a Grado  I-56010 Pisa (Italy), \break
\iRAL
  Rutherford Appleton Laboratory,
  Chiton,Didcot - Oxon OX11 0QX United Kingdom, \break
\iRUTG
  Rutgers University,
  Serin Physics Labs  Piscataway,NJ 08855-0849, \break
\iSLAC
  Stanford Linear Accelerator Center,
  2575 Sand Hill Road  Menlo Park,CA 94025, \break
\iSOGA
  Sogang University,
  Ricci Hall  Seoul, Korea, \break
\iSOONG
  Soongsil University,
  Dongjakgu Sangdo 5 dong 1-1    Seoul, Korea 156-743, \break
\iTENN
  University of Tennessee,
  401 A.H. Nielsen Physics Blg.  -  Knoxville,Tennessee 37996-1200, \break
\iTOHO
  Tohoku University,
  Bubble Chamber Lab. - Aramaki - Sendai 980 (Japan), \break
\iUCSB
  U.C. Santa Barbara,
  3019 Broida Hall  Santa Barbara,CA 93106, \break
\iUCSC
  U.C. Santa Cruz,
  Santa Cruz,CA 95064, \break
\iVAND
  Vanderbilt University,
  Stevenson Center,Room 5333  P.O.Box 1807,Station B  Nashville,TN 37235, \break
\iWASH
  University of Washington,
  Seattle,WA 98105, \break
\iWISC
  University of Wisconsin,
  1150 University Avenue  Madison,WS 53706, \break
\iYALE
  Yale University,
  5th Floor Gibbs Lab. - P.O.Box 208121 - New Haven,CT 06520-8121. \break

\rm
%

\end{center}


\vfill\eject

\vfill\eject

\renewcommand{\baselinestretch}{1.0}
 \begin{table}
 \footnotesize
\caption{
 Compilation of the systematic errors for the E0 algorithm and $y_{cut}=0.02$.
 The first column shows the error source, the second column the central value 
used, and the third column the variation considered.
  The remaining columns show the corresponding errors on the values of  
 $R_3^c/R_3^{uds}$ and $R_3^b/R_3^{uds}$; 
`+' (`$-$') denotes the error
corresponding to the relevant positive (negative) parameter variation.
}
 \begin{center}
 \begin{tabular}{|l||c|c|rr|rr|} \hline
   ~Source~  & ~Center~ & ~Variation~  &  \multicolumn{2}{c|} 
{$\delta R_3^c/R_3^{uds}$}  & \multicolumn{2}{c|} 
{$\delta R_3^b/R_3^{uds}$} \\ 
                             & ~Value~   &               & +     & $-$ & +  
   & $-$      \\ \hline
~tracking efficiency         & correction&     off       &   0.0020 &        & -0.0110 &\\
~2D imp. par. res.~          & smear     &     off       &  -0.0100 &        &  0.0080 & \\
~z track resolution~         & smear     &     off       &  0.0010 &         &  0.0120 & \\
~MC statistics               &0.8M       &      --       &  0.0190 & -0.0190 &  0.0091 & -0.0091 \\ \hline
~$B$ decay  $<n_{ch}>$       & $5.51$ trks & $\pm0.35$ trks & -0.0030 & -0.0026 &  0.0135 & -0.0132 \\
~$B$ fragm. $<x_b>$          & $0.697$   & $\pm0.008$    &$<$0.0001&  0.0004 &  0.0172 & -0.0191 \\
~$B$ fragm. shape            & Peterson  & Bowler        &  0.0021 &         & -0.0216 &          \\
~$B$ meson lifetime          & $1.56ps$  & $\pm0.05$ ps   & -0.0021 &  0.0022 & -0.0011 &  0.0009 \\
~$B$ baryon lifetime         & $1.10ps$  & $\pm0.08$ ps   & -0.0003 &  0.0003 &$<$0.0001& -0.0000 \\
~$B$ baryon prod.            & 7.6\%     &$\pm3.2\%$     &  0.0014 & -0.0016 &  0.0021 & -0.0023 \\
~$B \to D^+ + X$ fraction       & 0.192     & $\pm0.05$     &  0.0011 & -0.0012 & -0.0013 & -0.0008 \\
~\z0 \ra \bb: $f^b$          & 0.2156    & $\pm0.0017$   &  0.0022 & -0.0021 &  0.0014 & -0.0014 \\ \hline
~\z0 \ra \cc: $f^c$          & 0.172     & $\pm0.010$    &  0.0272 & -0.0294 &  0.0044 & -0.0042 \\
~$C$ fragm. $<x_c>$          & 0.483     & $\pm0.008$    &  0.0213 & -0.0211 &  0.0002 & -0.0002 \\
~$C$ fragm. shape            & Peterson  & Bowler          &  0.0042 &         &  0.0006 &            \\
~$D^0$ decay $<n_{ch}>$~     & 2.54 trks & $\pm0.06$ trks &  0.0044 & -0.0048 &  0.0006 & -0.0006 \\
~$D^+$ decay $<n_{ch}>$~     & 2.48 trks & $\pm0.06$ trks &  0.0069 & -0.0074 &  0.0012 & -0.0013 \\
~$D_s$ decay $<n_{ch}>$~     & 2.62 trks & $\pm0.31$ trks &  0.0039 & -0.0040 & -0.0004 &  0.0003 \\
~$D^0$ lifetime              & $0.418$ ps & $\pm0.004$ ps  & -0.0001 &  0.0001 & -0.0002 &  0.0001 \\
~$D^+$ lifetime              & $1.054$ ps & $\pm0.015$ ps  &  0.0001 & -0.0001 & -0.0001 &  0.0001 \\
~$D_s$ lifetime              & $0.466$ ps & $\pm0.017$ ps  &  0.0001 & -0.0001 & -0.0003 &  0.0003 \\
~$D^0 \to K^0$ mult.         & 0.402     & $\pm0.059$    &  0.0088 & -0.0089 &  0.0026 & -0.0026  \\
~$D^+ \to K^0$ mult.         & 0.644     & $\pm0.078$    &  0.0102 & -0.0120 &  0.0027 & -0.0027 \\
~$D_s \to K^0$ mult.         & 0.382     & $\pm0.057$    &  0.0012 & -0.0013 &  0.0003 & -0.0003 \\
~$D^0 \to$ no $\pi^0$ fraction & 0.370     & $\pm0.037$    &  0.0069 & -0.0075 &  0.0034 & -0.0034 \\
~$D^+ \to$ no $\pi^0$ fraction & 0.496     & $\pm0.050$    &  0.0017 & -0.0018 &  0.0029 & -0.0029 \\
~$D_s \to$ no $\pi^0$ fraction & 0.348     & $\pm0.035$    & -0.0002 &  0.0001 & -0.0003 &  0.0003 \\
~$c\bar c \to D^+ + X$ fraction & 0.259     & $\pm0.028$    &  0.0029 & -0.0034 &  0.0001 & -0.0002 \\
~$c\bar c \to D_s + X$ fraction & 0.113     & $\pm0.037$    & -0.0025 &  0.0019 &  0.0002 & -0.0002 \\
~$c\bar c \to \Lambda_c + X$ fraction & 0.074     & $\pm0.029$    & -0.0051 &  0.0044 & -0.0001 & -0.0001 \\
~$\Lambda_c$ decay $<n_{ch}>$~& 2.79      & $\pm0.45$ trks &  0.0003 & -0.0002 &  0.0024 & -0.0024 \\
~$\Lambda_c$ lifetime        & $0.216ps$ & $\pm0.011$ ps  & -0.0037 &  0.0011 & -0.0006 &  0.0001 \\
~$g \to bb$ rate             & $0.31$    & $\pm0.11\%$   &  0.0001 & -0.0001 & -0.0038 &  0.0039 \\
~$g \to cc$ rate             & $2.38$    & $\pm0.48\%$   & -0.0019 &  0.0020 & -0.0015 &  0.0016 \\
~$K^0$ prodn. & 0.658trks & $\pm0.050$ trks & -0.0051 &  0.0045 & -0.0061 &  0.0058 \\
~$\Lambda$ prodn. & 0.124trks & $\pm0.008$ trks & -0.0007 &  0.0009 & -0.0008 &  0.0009 \\ \hline
~Total Exp. Syst.            &           &               &  0.0440 & -0.0480 &  0.0300 & -0.0370 \\ \hline
~$Q_0$~                      & 1 GeV      & $^{+1}_{-0.5}$ GeV     &  0.0074 & -0.0027 &  0.0062 & -0.0237 \\
~$\sigma_q$~                 & 0.39 GeV   & $\pm0.04$ GeV  &  0.0042 & -0.0008 &  0.0015 &  0.0012 \\
~hadronisation model        & JETSET7.4 & HERWIG5.9     &  0.0123 &         & -0.0383 &  \\ \hline
 Total Hadronisation        &           &               &  0.0150 & -0.0028 &  0.0065 &  -0.0450 \\ \hline
\end{tabular}
\label{Table:syst}
\end{center}
\end{table}
\renewcommand{\baselinestretch}{1.4}

\clearpage

\begin{table}[htbp]
\caption{$R_3^i/R_3^{uds}$ and $\alpha_s^i/\alpha_s^{uds}$ values and errors.}
\vskip .5truecm
\label{Table:summary_option4}
\begin{center}
\begin{tabular}{|l|c|c|c|c|c|c|c|}  \hline
Algorithm           & E     &  E0   & P     & P0    & D     & G & \\ 
 $y_c$ & 0.040 & 0.020 & 0.020 & 0.015 & 0.010 & 0.080  & \\ \hline
 \multicolumn{7}{|c|} {$R_3^{c}/R_3^{uds}$ } & \\ \hline
central val.      & 1.043 & 1.066  & 1.004 &  1.058 & 1.038 & 1.040 & \\  \hline
 stat.     & 0.064 & 0.046  & 0.046 &  0.040 & 0.062 & 0.086 & \\ \hline
 exp. syst.     & $^{+0.065}_{-0.075}$ & $^{+0.044}_{-0.048}$  & 
$^{+0.046}_{-0.046}$ &  $^{+0.039}_{-0.039}$ & $^{+0.062}_{-0.067}$ & 
$^{+0.074}_{-0.085}$ & \\ \hline
 hadronisation      & $^{+0.012}_{-0.001}$ & $^{+0.015}_{-0.003}$ & 
$^{+0.014}_{-0.004}$ & $^{+0.015}_{-0.003}$ & $^{+0.006}_{-0.003}$ & 
$^{+0.008}_{-0.004}$ & \\ \hline
total. &  $^{+  0.092}_{ -0.099}$ & $^{+  0.065}_{ -0.067}$ & 
$^{+  0.067}_{ -0.065}$ & $^{+  0.058}_{ -0.056}$ & $^{+  0.088}_{ -0.091}$ & 
$^{+  0.114}_{ -0.121}$ & \\ \hline
 \multicolumn{7}{|c|} {$\alpha_s^{c}/\alpha_s^{uds}$ } & r.m.s \\ \hline
central val.      & 1.031& 1.054 &1.004 &1.052 &1.032 &1.035  & 0.017 \\ \hline
 stat.     &  0.046 &  0.037 &  0.041 &  0.035  & 0.051 &  0.074  & \\ \hline
   exp. syst. & $^{+  0.047}_{ -0.054}$ & $^{+  0.036}_{ -0.039}$ & 
$^{+  0.041}_{ -0.041}$ & $^{+  0.035}_{ -0.035}$ & 
$^{+  0.051}_{ -0.055}$ & $^{+  0.064}_{ -0.073}$ & \\ \hline
  hadronisation & $^{+  0.009}_{ -0.001}$ & $^{+  0.012}_{ -0.002}$ & 
$^{+  0.012}_{ -0.001}$ & $^{+  0.013}_{ -0.003}$ & $^{+  0.005}_{ -0.002}$ & 
$^{+  0.007}_{ -0.001}$ & \\ \hline
translation    &  $^{+0.001}_{-0.002}$ & $^{+0.005}_{-0.006}$ & $<$0.001 & 
$\pm0.008$ & $^{+0.003}_{-0.005}$ & $^{+0.004}_{-0.006}$ & \\ \hline
 \multicolumn{7}{|c|} {$R_3^{b}/R_3^{uds}$ }  & \\ \hline
central val.      & 1.050 & 1.054   & 1.048 & 1.055 & 0.964 & 0.995  & \\  \hline
 stat.     & 0.026 & 0.019  &  0.019 & 0.017 & 0.023 & 0.032 & \\ \hline
 exp. syst.     & $^{+0.038}_{-0.042}$ &  $^{+0.030}_{-0.037}$ & 
$^{+0.027}_{-0.037}$ & $^{+0.028}_{-0.035}$ & $^{+0.038}_{-0.041}$ & 
$^{+0.035}_{-0.036}$ & \\ \hline
 hadronisation     &  $^{+0.011}_{-0.046}$ & $^{+0.007}_{-0.045}$ & 
$^{+0.002}_{-0.026}$ & $^{+0.007}_{-0.037}$ & $^{+0.001}_{-0.006}$ & 
$^{+0.020}_{-0.008}$ & \\ \hline
total. &  $^{+  0.047}_{ -0.067}$ & $^{+  0.036}_{ -0.061}$ & 
$^{+  0.033}_{ -0.049}$ & $^{+  0.033}_{ -0.054}$ & $^{+  0.044}_{ -0.047}$ & 
$^{+  0.051}_{ -0.049}$ & \\ \hline
 \multicolumn{7}{|c|} {$\alpha_s^{b}/\alpha_s^{uds}$ } & r.m.s \\ \hline
central val.     & 0.989 & 0.995 & 1.018 & 1.014 & 1.009 & 0.993 & 0.011 \\ \hline
 stat.     & 0.018 & 0.015 & 0.017 & 0.015 & 0.021 & 0.027 & \\ \hline
    exp. syst. & $^{+  0.026}_{ -0.029}$ & $^{+  0.023}_{ -0.028}$ & 
$^{+  0.023}_{ -0.032}$ & $^{+  0.024}_{ -0.030}$ & $^{+  0.034}_{ -0.036}$ & 
$^{+  0.030}_{ -0.031}$ & \\  \hline
 hadronisation & $^{+  0.008}_{ -0.032}$ & $^{+  0.005}_{ -0.034}$ & 
$^{+  0.002}_{ -0.022}$ & $^{+  0.006}_{ -0.032}$ & $^{+  0.001}_{ -0.005}$ & 
$^{+  0.017}_{ -0.007}$ & \\ \hline
translation     &   $^{+0.016}_{-0.015}$ & $^{+0.013}_{-0.015}$ &  
$^{+0.011}_{-0.014}$ &  $^{+0.017}_{-0.018}$ &  $\pm0.012$ &  
$^{+0.008}_{-0.009}$  & \\ \hline
\end{tabular}
\end{center}
\end{table}

\clearpage

\begin{table}[htbp]
\footnotesize
\begin{center}
\caption{The coefficients $A^b, B^b, C^b$ for the Next-to-Leading-Order
 calculation for massive quarks. The numbers in parentheses represent the 
estimated numerical precision. Theoretical uncertainties in the computation of
the $B^b$ coefficients derive from the `slicing parameter' used to isolate
singular regions of phase space, as well as from the conversion to the
$\overline{MS}$ quark mass parameter
Effects from 
higher-order perturbative QCD contributions are discussed in the text. }

\vskip .5truecm

\label{Table:Aachen}
\begin{tabular}{|l|r||r|r|r||r|r|r|r|} \hline \ & \  &
 \multicolumn{3}{c||}{$A^b$ for $m_b(M_{Z^0})$ (GeV/$c^2$) =} &  
 \multicolumn{3}{c|}{$B^b$ for $m_b(M_{Z^0})$ (GeV/$c^2$) =} \\ \hline
Algorithm  & $y_c$ & 2.5  & 3  & 3.5  & 2.5  & 3   & 3.5  \\  \hline
E & 0.040 &  14.392(1) & 14.459(1) & 14.543(1) & 443(4) & 466(4) & 487(4)  \\ \hline
E0& 0.020 &  24.850(2) & 25.024(2) & 25.231(2) & 277(4) & 291(4) & 310(4) \\ \hline
P & 0.020 &  24.850(2) & 25.024(2) & 25.231(2) & 63(4)  &67(4)   & 75(4)  \\ \hline
P0& 0.015 &  30.054(2) & 30.315(2) & 30.631(2) & 2(4)   & 14(4)  & 29(4) \\ \hline
D & 0.010 &  15.355(2) & 15.213(2) & 15.060(2) & 105(4) & 102(4) & 99(4) \\ \hline
G & 0.080 &  11.493(1) & 11.435(1) & 11.365(1) & 61(4)  & 58(4)  & 57(4) \\ \hline
\end{tabular}

\vskip 1truecm

\begin{tabular}{|l|r||r|r|r|r|} \hline
\ & \  & \multicolumn{3}{c|}{$C^b$ for $m_b(M_{Z^0})$ (GeV/$c^2$) =} \\ \hline
Algorithm & $y_c$ &  2.5  & 3 & 3.5  \\ \hline
E & 0.040 &  27.91(1)  & 28.27(1) & 28.71(1) \\ \hline
E0& 0.020 &  125.39(7) & 127.34(7)&129.55(8) \\ \hline
P & 0.020 &  125.39(7) & 127.34(7)&129.55(8) \\ \hline
P0& 0.015 &  202.8(1)  & 206.1(1) & 209.4(1) \\ \hline
D & 0.010 &  84.30(6)  & 82.83(6) & 81.19(6) \\ \hline
G & 0.080 &  65.55(4)  & 64.60(3) & 63.56(3) \\ \hline
\end{tabular}
\end{center}
\end{table}

\clearpage

\renewcommand{\baselinestretch}{1.0}
 \begin{table}[htbp]
\footnotesize
 \caption{Summary of translation uncertainties on the \alp ratios
for each algorithm; `+' (`$-$') denotes the error
corresponding to the relevant positive (negative) parameter variation.
}
\vskip .5truecm
 \label{Table:translation}
\begin{center}
\begin{tabular}{|l|c|c|rr|rr|} \hline
   ~source~  & Center & Variation & \multicolumn{2}{c|} {$\delta 
\alpha_s^c/\alpha_s^{uds}$ (\%)}  & \multicolumn{2}{c|} {$\delta 
\alpha_s^b/\alpha_s^{uds}$ (\%) } \\
 & & & + & $-$ & + & $-$  \\ \hline
  &  \multicolumn{6}{|c|}{E-algo ($y_c=0.04$)} \\ \hline
 $m_b(M_Z)$ & 3.0GeV & $\pm0.5$    & 0.000 & 0.000 & -0.014 &  0.015 \\ \hline
 $\mu, \alpha_s$ dep.&        &    & -0.002 & 0.001 & 0.006 & -0.006 \\ \hline
 $\geq 4$jet contrib. &$C$ &$\pm C$ & $<$0.001 & $<$0.001 & -0.001 & 0.001 \\ 
\hline
 Total             &      &        &  0.001 & -0.002 & 0.016 & -0.015 \\ 
\hline \hline

  &  \multicolumn{6}{|c|}{E0-algo ($y_c=0.02$)} \\ \hline 
 $m_b(M_Z)$ & 3.0GeV & $\pm0.5$    & 0.000 & 0.000 &-0.014 & 0.012 \\ \hline
 $\mu, \alpha_s$ dep.&        &    & -0.005 & 0.004 & 0.005 & -0.005 \\ \hline
 $\geq 4$jet contrib.&$C$ &$\pm C$ &  -0.002 & 0.003 & -0.001 &  0.002 \\  
\hline
 Total             &      &        & 0.005 & -0.006 & 0.013 & -0.015 \\  
\hline \hline

  &  \multicolumn{6}{|c|}{P-algo ($y_c=0.02$)} \\ \hline
 $m_b(M_Z)$ & 3.0GeV & $\pm0.5$    & 0.000 & 0.000 &-0.012 & 0.009 \\ \hline
 $\mu, \alpha_s$ dep.&        &    & $<$0.001 & $<$0.001 & -0.002 &  0.002 \\  
\hline
 $\geq 4$jet contrib.&$C$ &$\pm C$ & $<$0.001 & $<$0.001 & -0.005 & 0.007 \\  
\hline
 Total             &      &        & $<$0.001 & $<$0.001 & 0.011 & -0.014 \\ 
\hline \hline

  &  \multicolumn{6}{|c|}{P0-algo ($y_c=0.015$)} \\ \hline
 $m_b(M_Z)$ & 3.0GeV & $\pm0.5$    & 0.000 & 0.000 &-0.017 & 0.015 \\ \hline
 $\mu, \alpha_s$ dep.&        &    & -0.007 & 0.005 & -0.001 & $<$0.000 \\ 
\hline
 $\geq 4$jet contrib.&$C$ &$\pm C$ & -0.004 & 0.006 & -0.006 & 0.008 \\ \hline
 Total             &      &        & 0.008 & -0.008 & 0.017 & -0.018 \\ 
\hline \hline

  &  \multicolumn{6}{|c|}{D-algo ($y_c=0.010$)} \\ \hline
 $m_b(M_Z)$ & 3.0GeV & $\pm0.5$    & 0.000 & 0.000 & 0.011 &-0.010 \\ \hline
 $\mu, \alpha_s$ dep.&        &    & -0.005 & 0.002 &-0.005 & 0.003 \\ \hline
 $\geq 4$jet contrib.&$C$ &$\pm C$ & -0.002 & 0.002 & -0.003 & 0.003 \\  \hline
 Total             &      &        & 0.003 & -0.005 & 0.012 &-0.012 \\  
\hline \hline

  &  \multicolumn{6}{|c|}{G-algo ($y_c=0.08$)} \\ \hline
 $m_b(M_Z)$ & 3.0GeV & $\pm0.5$    & 0.000 & 0.000 & 0.010 &-0.009 \\ \hline
 $\mu, \alpha_s$ dep.&        &    & -0.005 & 0.003 & 0.005 & -0.003 \\ \hline
 $\geq 4$jet contrib.&$C$ &$\pm C$ & -0.002 & 0.003 & 0.001 & -0.001 \\  \hline
 Total             &      &        & 0.004 &-0.006 & 0.008 &-0.009 \\  \hline 
 \end{tabular}
\end{center}
 \end{table}
\renewcommand{\baselinestretch}{1.4}

\vfill
\eject

\begin{figure}[htbp]
 \begin{center}
 \leavevmode
 \epsfysize=6in
 \epsfbox{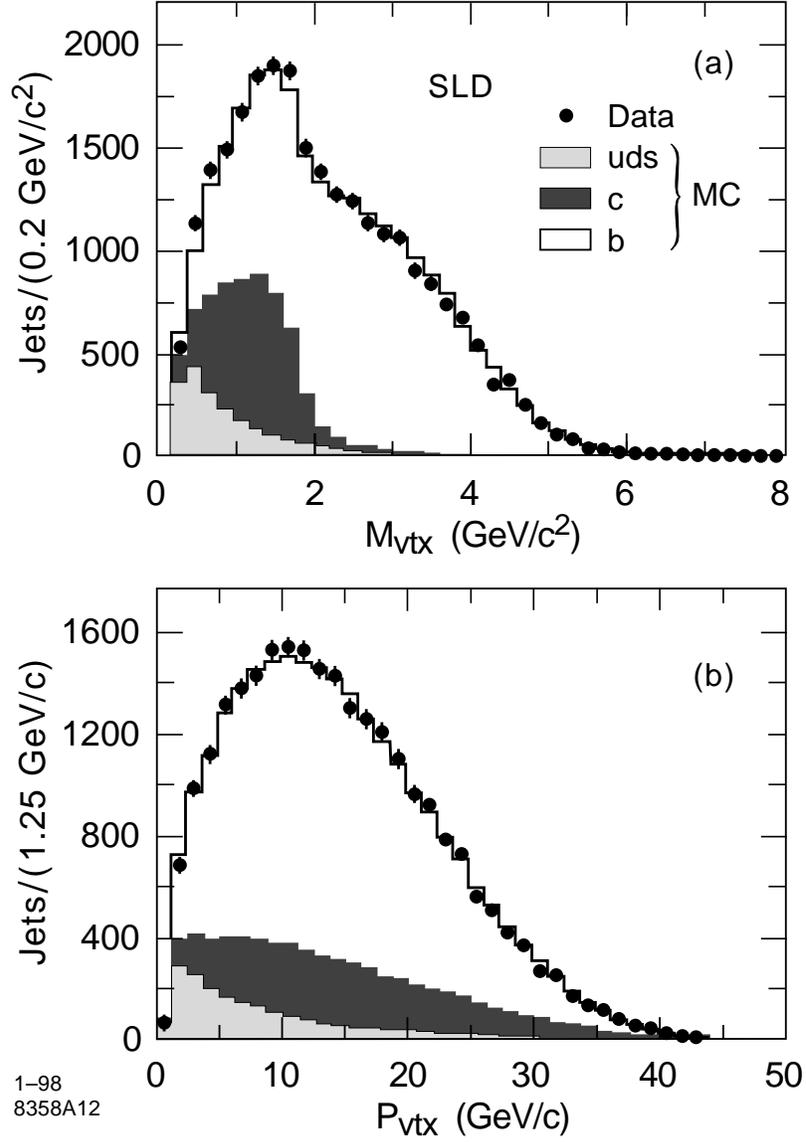} 
 \end{center}
\caption{The distributions of (a) the vertex mass, $M_{vtx}$, and (b) 
the vertex momentum, $P_{vtx}$, in our data sample (points);
the simulated distributions are shown as a histogram in which the contributions
from events of different primary quark flavor are indicated.
}
\end{figure}

\begin{figure}[htbp]
 \begin{center}
 \leavevmode
 \epsfysize=4in
 \epsfbox{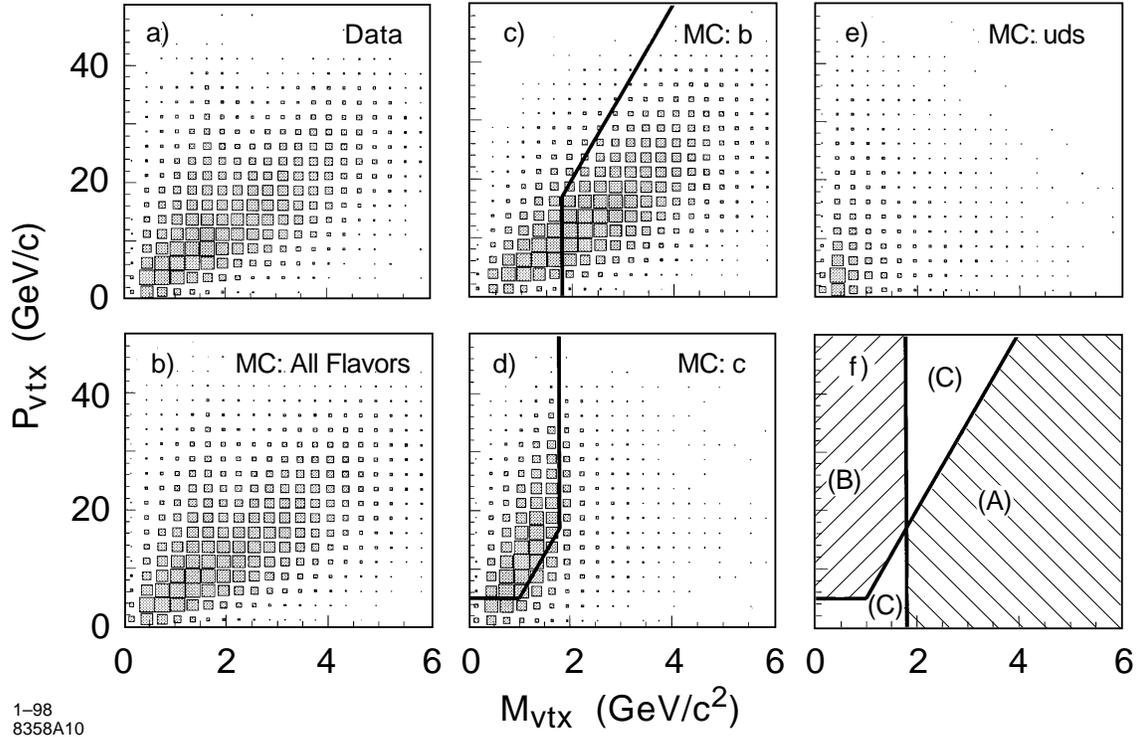} 
 \end{center}
\caption{The two-dimensional distribution of vertex momentum $P_{vtx}$ vs. 
vertex mass $M_{vtx}$ (see text).
(a) Data; (b) all-flavors simulation; (c) \bb event simulation;
(d) \cc event simulation; (e) $q_l\overline{q_l}$ simulation. In (f) 
the regions used for
$b$-tagging (A), $c$-tagging (B) and no-tagging (C) are indicated (see text).
}
\end{figure}

\begin{figure}[htbp]
 \begin{center}
 \leavevmode
 \epsfysize=4in
 \epsfbox{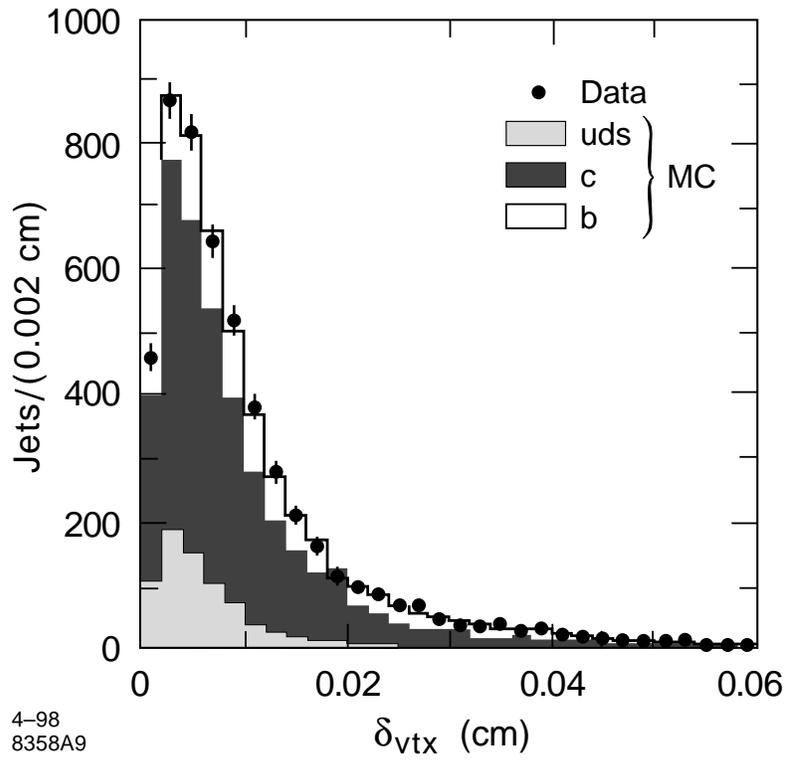} 
 \end{center}
\caption{The distribution of vertex impact parameter, $\delta_{vtx}$, for 
events containing vertices in region (B): data (points);
the simulated distribution is shown as a histogram in which the contributions
from events of different primary quark flavor are indicated.
}
\end{figure}

\begin{figure}[htbp]
 \begin{center}
 \leavevmode
 \epsfysize=4in
 \epsfbox{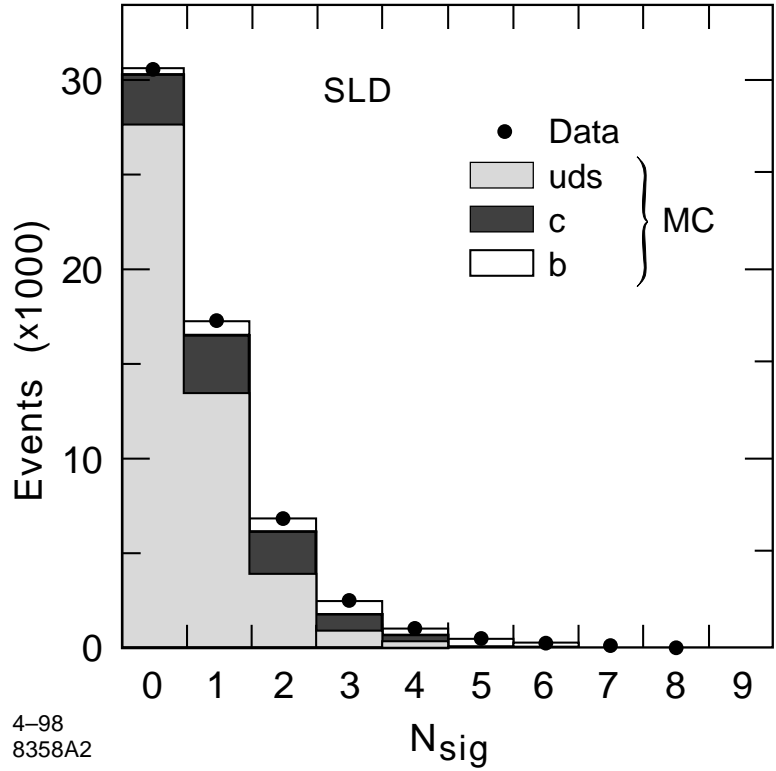} 
 \end{center}
\caption{The distribution of the number of tracks per event that miss the IP
by at least 
2$\sigma$ in terms of their impact parameter in the plane normal to the
beamline, in events that contain no reconstructed vertex (see text); data 
(points). The simulated distribution is shown as a histogram in which the 
contributions from events of different primary quark flavor are indicated.
}
\end{figure}

\begin{figure}[htbp]
 \begin{center}
 \leavevmode
 \epsfysize=5in
 \epsfbox{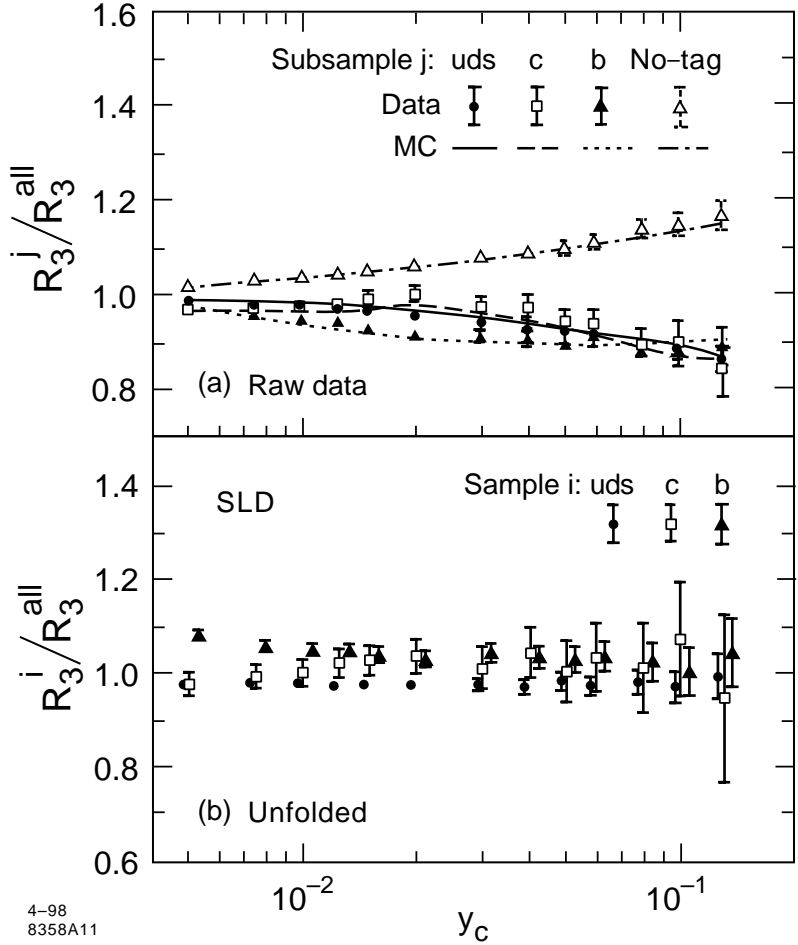} 
 \end{center}
\caption{(a) The raw measured ratios \rj, $1\leq j\leq 4$, vs. $y_c$ for the 4 
subsamples 
(see text); data (points with error bars), and simulation (lines joining values
at the same $y_c$ values as the data). 
(b) The unfolded ratios \ri, $i=b,c,uds$, vs. $y_c$ for the 3 primary
event flavor groups. Only statistical errors are shown. In (b) points
corresponding to a common $y_c$ value have been displaced horizontally for
clarity.
}
\end{figure}

\begin{figure}[htbp]
 \begin{center}
 \leavevmode
 \epsfysize=4in
 \epsfbox{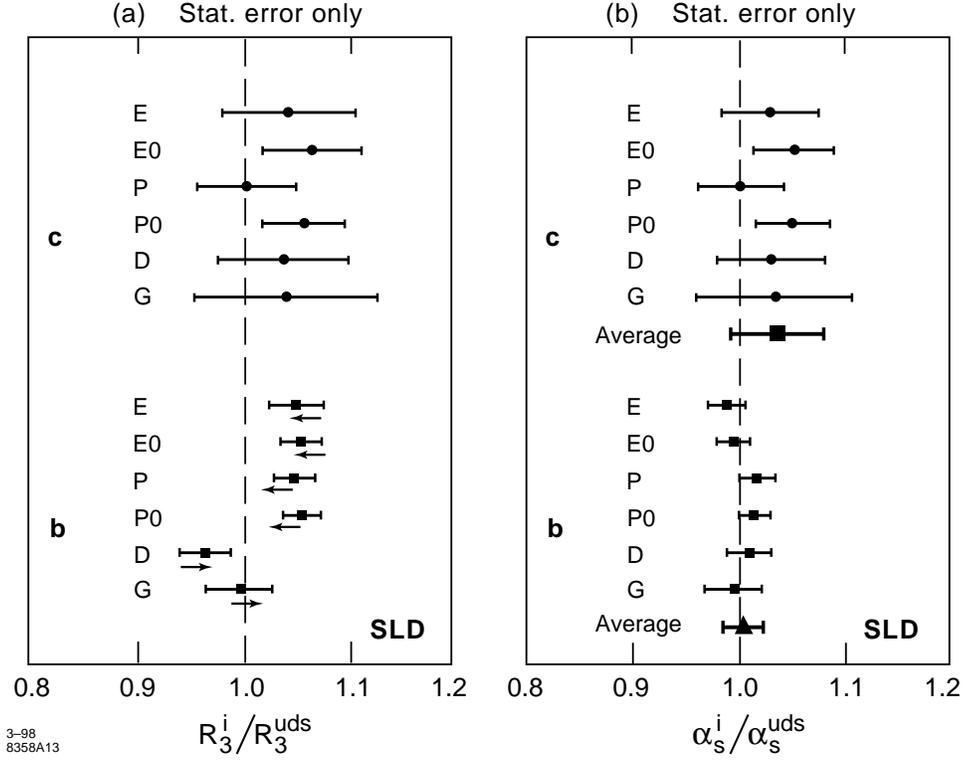} 
 \end{center}
\caption{
(a) The measured ratios $R_3^i/R_3^{uds}$, and (b)
 the corresponding translated ratios $\alpha_s^i/\alpha^{uds}$ ($i=c, b$).
The arrows in (a) indicate the range of the theoretical prediction described 
in the text for
values of the $b$-quark mass in the range $2.5\leq m_b(M_{Z^0})\leq 3.5$ 
GeV/$c^2$, with the arrow pointing towards the lower mass value. 
In (b) The weighted average over the six algorithms is also shown. In all cases
only statistical error bars are displayed.}
\label{Fig:results}
\end{figure}

\begin{figure}[htbp]
 \begin{center}
 \leavevmode
 \epsfysize=4in
 \epsfbox{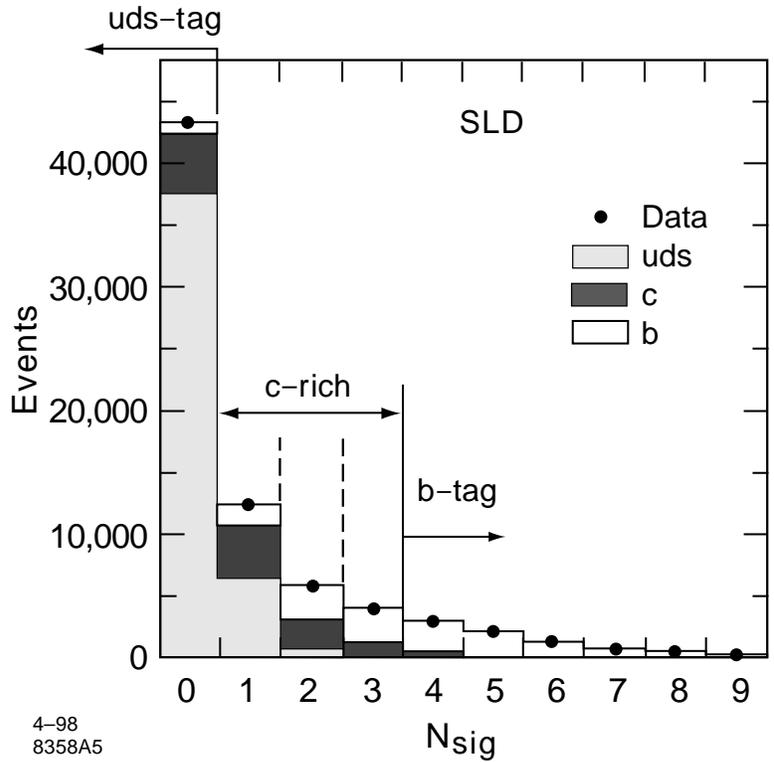} 
 \end{center}
\caption{The distribution of the number of tracks that miss the IP by at least 
3$\sigma$ in terms of their impact parameter in the plane normal to the
beamline (see text): data (points);
the simulated distribution is shown as a histogram in which the contributions
from events of different primary quark flavor are indicated.
}
\end{figure}

\begin{figure}[htbp]
 \begin{center}
 \leavevmode
 \epsfysize=5in
 \epsfbox{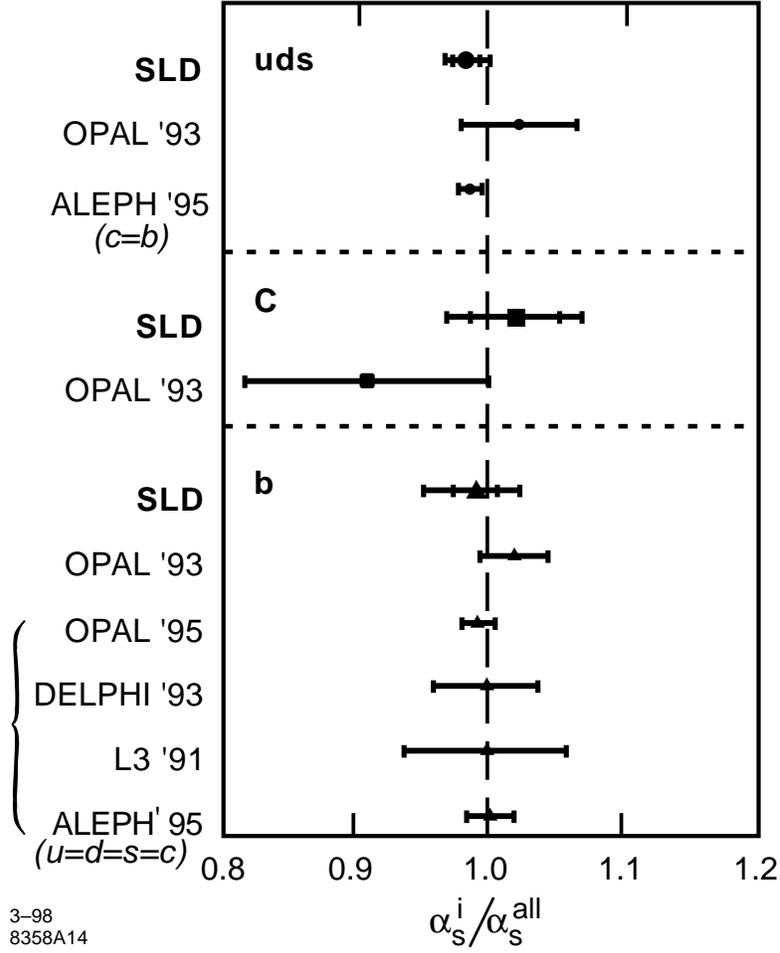} 
 \end{center}
\caption{
Summary of measurements of $\alpha_s^i/\alpha_s^{all}$ ($i$ = $uds$, $c$ or 
$b$) from experiments at the
\z0  resonance. We derived the ALEPH $\alpha_s^{uds}/\alpha_s^{all}$ value 
from their measured value of $\alpha_s^{uds}/\alpha_s^{bc}$, 
as well as the five
bracketed LEP values of $\alpha_s^{b}/\alpha_s^{all}$ from the measured values
of  $\alpha_s^{b}/\alpha_s^{udsc}$,
by assuming $\alpha_s^{all}$ = $\sum_{uds,c,b} f^i\, 
\alpha_s^i$, where
$f^i$ is the Standard Model branching fraction for \z0 decays to quark flavor
$i$.
}
\end{figure}
 
\end{document}